%% file: main.tex
\documentclass[11pt, a4paper, logo, copyright]{googledeepmind}

\pdftrailerid{redacted}

\makeatletter
\renewcommand\bibentry[1]{\nocite{#1}{\frenchspacing\@nameuse{BR@r@#1\@extra@b@citeb}}}
\makeatother

\usepackage{kantlipsum, lipsum}
\usepackage{dsfont}
\usepackage{gdm-colors}
\usepackage[utf8]{inputenc}   
\usepackage{newunicodechar}   
\usepackage{wasysym,marvosym}
\usepackage{ulem}

\usepackage{tikz}
\usetikzlibrary{shapes.geometric, arrows.meta, positioning}
\usepackage{amsmath}
\usepackage{algorithm}
\usepackage{algpseudocode}
\usepackage{listings}
\usepackage{enumitem} 
\usepackage{lipsum}
\usepackage[most]{tcolorbox}
\definecolor{thinkcolor}{RGB}{227,196,144}
\definecolor{observecolor}{RGB}{153,201,227}
\definecolor{explorecolor}{RGB}{178,217,200}

\tcbset{
    common/.style={
		enhanced,
		arc=0mm,
		fonttitle=\large\bfseries,
		coltitle=black,
		attach boxed title to top left={xshift=0mm,
										yshift=-0.50mm},
		boxed title style={
			skin=enhancedfirst jigsaw,
			size=small,
			arc=5mm,
			bottom=0mm,
			left=8mm,
			right=18mm,
			top=1mm},
			boxrule=0pt,
			frame hidden},
    thinkstyle/.style={
		common,
		colbacktitle=thinkcolor,
		colframe=thinkcolor,
		colback=thinkcolor!40,
		borderline north={4pt}{0pt}{thinkcolor}},
	observestyle/.style={
		common,
		colbacktitle=observecolor,
		colframe=observecolor,
		colback=observecolor!40,
		borderline north={4pt}{0pt}{observecolor}}
}

\newtcolorbox{think}{thinkstyle,title=Prompt Template}
\newtcolorbox{observe}{observestyle,title=observe}
\newtcolorbox{custom}[2][gray]{
	common,
	title=#2,
	colbacktitle=#1,
	colframe=#1,
	colback=#1!40,
	borderline north={4pt}{0pt}{#1}}
\usepackage{array, multirow, tabularx, booktabs, makecell}

\definecolor{interest_colframe}{rgb}{0.8, 0.878, 0.871}
\definecolor{interest_colback}{rgb}{0.918, 0.953, 0.949}

\definecolor{tag_colframe}{rgb}{0.965, 0.898, 0.847}
\definecolor{tag_colback}{rgb}{0.988, 0.961, 0.941}

\definecolor{exp_colback}{rgb}{0.949, 0.965, 0.980}
\definecolor{exp_colframe}{rgb}{0.878, 0.922, 0.965}

\tcbset{
    promptbox/.code args={#1/#2}{
        \tcbset{
            enhanced,
            arc=0mm,
            colframe=#1, 
            colback=#2, 
            coltitle=black,
            fonttitle=\large\bfseries,
            attach boxed title to top left={xshift=0mm, yshift=-1.0mm},
            boxed title style={
                skin=enhancedfirst jigsaw,
                size=small,
                arc=3mm,
                bottom=0mm,
                left=8mm,
                right=8mm,
                top=1mm,
                colback=#1
            },
            boxrule=0pt,
            frame hidden,
            borderline north={4pt}{0pt}{#1},
        }
    }
}

\newcommand{\assignmentQuestionName}{Question} 

\usepackage{caption}
\usepackage{booktabs}
\usepackage{arydshln}
\usepackage{dashrule}

\usepackage[authoryear, sort&compress, round]{natbib}

\usepackage{bbding}
\usepackage[T1]{fontenc}    
\usepackage{hyperref}       
\usepackage{url}            
\usepackage{booktabs}       
\usepackage{nicefrac}       
\usepackage{microtype}      
\usepackage{amsmath}
\usepackage{graphicx}
\usepackage{multicol}
\usepackage[nameinlink]{cleveref}
\usepackage{bbm}
\usepackage{multirow}
\usepackage{soul}
\usepackage{float}
\usepackage{wrapfig}
\usepackage{blindtext}
\usepackage{tablefootnote}
\usepackage{amsfonts}
\usepackage[flushleft]{threeparttable}
\usepackage{colortbl}
\usepackage{mathtools}
\usepackage{bm}
\usepackage{makecell}
\usepackage{caption}
\usepackage{capt-of}
\usepackage{array}
\usepackage{calc}      
\usepackage{caption}   
\usepackage{subcaption}  
\usepackage[bottom]{footmisc}
\usepackage{fontawesome}
\usepackage{tabularx}        
\usepackage{siunitx}

\newcolumntype{C}{>{\centering\arraybackslash}X}
\newcolumntype{L}{>{\raggedright\arraybackslash}X}

\title{OnePiece: Bringing Context Engineering and Reasoning to Industrial Cascade Ranking System}

\author[1]{Sunhao Dai\textsuperscript{\dag}}
\author[1]{Jiakai Tang\textsuperscript{\ddag}}
\author[2]{Jiahua Wu}
\author[2]{Kun Wang}
\author[2]{Yuxuan Zhu}
\author[2]{Bingjun Chen}
\author[2]{Bangyang Hong}
\author[2]{Yu Zhao}
\author[2]{Cong Fu\textsuperscript{\P}}
\author[2]{Kangle Wu}
\author[2]{Yabo Ni}
\author[2]{Anxiang Zeng}
\author[3]{Wenjie Wang}
\author[1]{Xu Chen}
\author[1]{Jun Xu}
\author[4]{See-Kiong Ng}

\affil[1]{Renmin University of China}
\affil[2]{Shopee}
\affil[3]{University of Science and Technology of China}
\affil[4]{National University of Singapore}

\setauthemails{\textsuperscript{\dag}sunhaodai@ruc.edu.cn, \textsuperscript{\ddag}tangjiakai5704@ruc.edu.cn,\textsuperscript{\P}fc731097343@gmail.com}

\begin{abstract}
Despite the growing interest in replicating the scaled success of large language models (LLMs) in industrial search and recommender systems, most existing industrial efforts remain limited to transplanting Transformer architectures, which bring only incremental improvements over strong Deep Learning Recommendation Models (DLRMs).
From a first principle perspective, the breakthroughs of LLMs stem not only from their architectures but also from two complementary mechanisms: \textit{context engineering}, which enriches raw \textit{input} queries with contextual cues to better elicit model capabilities, and \textit{multi-step reasoning}, which iteratively refines model \textit{outputs} through intermediate reasoning paths. However, these two mechanisms and their potential to unlock substantial improvements remain largely underexplored in industrial ranking systems.

In this paper, we propose OnePiece, a unified framework that seamlessly integrates LLM-style context engineering and reasoning into both retrieval and ranking models of industrial cascaded pipelines. OnePiece is built on a pure Transformer backbone and further introduces three key innovations: (1) structured context engineering, which augments interaction history with preference and scenario signals and unifies them into a structured tokenized input sequence for both retrieval and ranking; (2) block-wise latent reasoning, which equips the model with multi-step refinement of representations and scales reasoning bandwidth via block size; (3) progressive multi-task training, which leverages user feedback chains to effectively supervise reasoning steps during training.
Extensive offline experiments verify the effectiveness of each core design and show that OnePiece not only achieves higher sample efficiency but also continues to benefit from larger training spans, surpassing strong baselines with fewer days of logs and maintaining improvement as more data becomes available.
OnePiece has been deployed in the main personalized search scenario of Shopee and achieves consistent online gains across different key business metrics, including over $+2\%$ GMV/UU and a $+2.90\%$ increase in advertising revenue. 
\end{abstract}

\begin{document}
\maketitle

\begin{figure}[H]
    \centering
    \includegraphics[width=\linewidth]{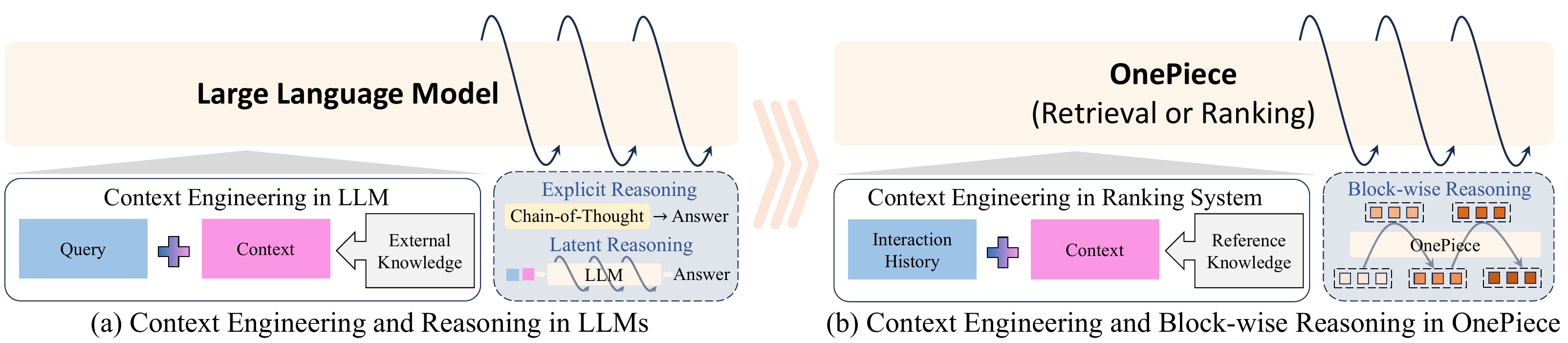}
    \caption{Bringing LLM mechanisms of context engineering and reasoning to industrial ranking systems. (a) In LLMs, context engineering broadens the query with external knowledge from the input side, while explicit or latent reasoning guides the output process, together enhancing the model’s ability to generate accurate answers. (b) In OnePiece, user interaction history is augmented with additional reference signals, while block-wise reasoning progressively refines representations, jointly improving performance in both retrieval and ranking modes.}
    \label{fig:intro}
\end{figure}

\newpage
\setcounter{tocdepth}{2} 

\tableofcontents 

\newpage

\input{section/intro}
\input{section/prelinminary}

\input{section/method}
\input{section/experiment}
\input{section/related_work}

\section{Conclusion and Future Directions}
In this paper, we introduce \textbf{OnePiece}, a unified framework that brings structured context engineering and block-wise latent reasoning into industrial cascaded ranking systems. 
Built upon a pure Transformer backbone, OnePiece incorporates context engineering to organize heterogeneous signals (interaction history, preference anchors, situational descriptors, and candidate items) into a structured tokenized sequence, equips the model with multi-step reasoning capacity, and optimizes this process through a progressive multi-task training strategy that leverages naturally available user feedback chains.
Extensive offline experiments validate the effectiveness of each design, reveal favorable scaling with respect to preference anchor length, training data span, and block size.
Online A/B testing at Shopee’s main personalized search scenario demonstrates that OnePiece achieves consistent gains across diverse business metrics, including over $+2\%$ GMV/UU improvement and $+2.90\%$ advertising revenue improvement, while also exhibiting superior efficiency and hardware utilization. 
These results position OnePiece as a promising new paradigm for building scalable, reasoning-driven ranking models in real-world industrial environments.

\begin{figure}[t]
    \centering
    \includegraphics[width=\linewidth]{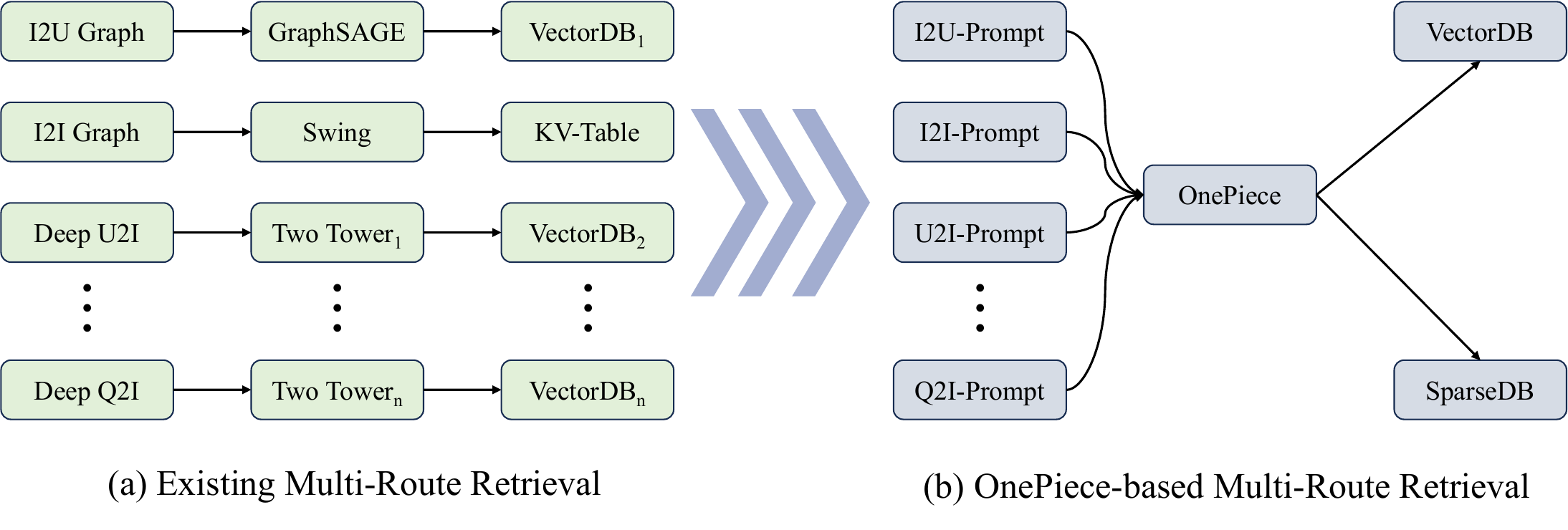}
    \caption{Comparison between existing multi-route retrieval systems and OnePiece-based unified architecture. (a) Traditional multi-route retrieval requires maintaining separate models with distinct parameters for different retrieval pathways (I2U, I2I, U2I, Q2I, etc.), each utilizing specialized architectures and storage systems. (b) OnePiece achieves unified multi-route retrieval through a single model that processes tailored prompts for different retrieval scenarios, enabling ``One For All" functionality while reducing system complexity and maintenance overhead.}
    \label{fig:futurework_ce}
\end{figure}

Looking ahead, several promising research directions emerge from this work:
\begin{itemize}
\item \textbf{Unified Multi-Route Retrieval:} As illustrated in Fig.~\ref{fig:futurework_ce}, existing multi-route retrieval systems require distinct design principles, training data, and model parameters for each retrieval pathway to achieve diversified personalization and comprehensive coverage of user interests. However, maintaining multiple heterogeneous models is resource-intensive and prone to redundancy across different routes. Our unified OnePiece architecture paves the way for ``\textbf{One For All}'' multi-route retrieval, where a single unified model can serve diverse recommendation objectives through tailored context engineering (e.g., adapting interaction history, preference anchors, and situational descriptors based on specific scenario characteristics).
The empirical evidence from Sec.~\ref{sec:recall_coverage} demonstrates strong diversity and exclusivity performance, supporting the feasibility of streamlined multi-route system design in industrial applications.
\item \textbf{Scalable Latent Reasoning:} While this work represents the first successful deployment of latent reasoning in industrial-scale ranking systems through multi-task progressive supervision, this approach also reveals inherent limitations in reasoning scalability. The challenge lies in obtaining sufficient multi-task signals to effectively supervise intermediate reasoning processes, which constrains our ability to scale reasoning capabilities further. Future research should explore more effective scaling methodologies for latent reasoning, such as incorporating online user feedback through reinforcement learning to adaptively determine optimal reasoning depth, or developing organic integration between model self-exploration and multi-task supervision processes that enable autonomous reasoning evolution.
\end{itemize}

\addcontentsline{toc}{section}{References}
\bibliographystyle{abbrvnat}
\nobibliography*
\bibliography{reference}

\clearpage

\appendix
\section*{Appendix}
\input{section/appendix}

\end{document}

%% file: section/intro.tex
\section{Introduction}

Large Language Models (LLMs), particularly those equipped with reasoning capabilities, have achieved remarkable success across a wide range of complex natural language processing (NLP) tasks~\citep{zhao2023survey}. Inspired by these advances, researchers have begun exploring how to replicate the key mechanisms behind LLM success in other domains in the hope of realizing similar breakthroughs~\citep{zhu2023large, lin2025can}. In industrial ranking systems (\textit{e.g.}, search and recommender systems), most efforts have focused on transplanting Transformer-based sequential architectures~\citep{wu2024survey}. While these models bring certain benefits, the improvements over strong Deep Learning Recommendation Models (DLRMs)~\citep{naumov2019deep} remain incremental, as core mechanisms such as attention and sequential feature modeling are already deeply integrated into classical designs.

Beyond their foundational Transformer-based architectures, the breakthroughs of LLMs can also be traced to two complementary mechanisms operating at different stages of the modeling process:
(1) \textbf{Context Engineering}~\citep{dong2024survey, mei2025survey}, which strengthens the \textit{input side} by enriching raw queries with structured contextual cues to better elicit model capacity; and
(2) \textbf{Multi-Step Reasoning}~\citep{huang2023towards, zhu2025survey}, which enhances the \textit{output side} by refining predictions iteratively through intermediate latent or explicit reasoning steps. From a first principle view, both context engineering and multi-step reasoning serve the same ultimate goal of eliciting stronger model capacity, one from the \textit{input} side and the other from the \textit{output} side.
Together, these two mechanisms substantially expand the generalization ability and capability frontier of LLMs, yet they remain largely underexplored in industrial ranking systems.

However, transplanting these two mechanisms directly into industrial ranking systems is far from straightforward. Unlike LLMs, ranking models cannot readily exploit prompt-style contexts or chain-of-thought supervision, making it unclear how to effectively operationalize context engineering and reasoning in this domain. Specifically, bringing these mechanisms into ranking requires us to tackle two fundamental challenges:
\vspace{-0.3cm}
\begin{itemize}[leftmargin=0.5cm]
\item \textbf{How to construct an informative input context?} Most transformer-based industrial models rely primarily on raw user–item interaction sequences, which lack the structural richness of LLM-style prompts~\citep{zhai2024actions, chen2025onesearch, deng2025onerec}. Moreover, existing feature engineering practices remain predominantly tailored to DLRM-style architectures, leaving open the question of how to enrich context to better endow ranking models with reasoning capabilities.
\item \textbf{How to optimize multi-step reasoning?} In LLMs, large-scale chain-of-thought annotations provide explicit guidance for training and optimizing reasoning processes~\citep{hao2024training, shen2025codi}. In contrast, industrial ranking systems lack such supervision, and even domain experts cannot feasibly articulate the latent decision paths underlying user behaviors in natural language~\citep{tang2025think}. This makes it difficult to directly supervise reasoning trajectories.
\end{itemize}
\vspace{-0.2cm}

To bridge this gap, we propose OnePiece, a unified framework that seamlessly integrates LLM-style context engineering and reasoning into both the retrieval and ranking stages of industrial cascaded pipelines.
\textbf{First}, OnePiece unifies input representation across retrieval and ranking through \textit{structured context engineering}. Specifically, it enriches user interaction history with preference anchors (auxiliary item sequences constructed from domain knowledge provide reference signals, \textit{e.g.}, top-clicked items under the current query) and situational descriptors (\textit{e.g.}, user profiles and query contextual features). For ranking, OnePiece further augments the context with a compact candidate item set, where items are jointly visible to each other, enabling the model to capture cross-candidate interactions for more accurate score prediction.
\textbf{Second}, OnePiece introduces \textit{block-wise latent reasoning}, where hidden states are progressively refined across multiple reasoning blocks, each building upon the last, thereby expanding the reasoning bandwidth and yielding more expressive representations.  \textbf{Finally}, to effectively supervise this multi-step process, OnePiece adopts a \textit{progressive multi-task training} strategy that leverages naturally available user feedback chains (\textit{e.g.}, click, add-to-cart, order) as staged supervision signals. This design provides structured guidance for intermediate reasoning steps and enables the model to gradually align with deeper levels of user preference, from surface engagement to final conversion.

We evaluate the effectiveness of OnePiece through extensive offline experiments and online A/B tests. A series of ablation studies demonstrates the effectiveness of each key module: context engineering improves Recall@100 by $+0.110$ and click-AUC by $+0.109$, while block-wise reasoning adds another $+0.047$ Recall@100 and $+0.030$ click-AUC. 
Moreover, OnePiece achieves stronger data efficiency than DLRM and continues to improve as training spans grow longer, while DLRM quickly plateaus.
Beyond offline evaluation, OnePiece has been fully deployed in Shopee’s main personalized search scenario, serving billions of users. Online A/B testing confirms that OnePiece delivers consistent business gains: retrieval raises GMV/UU by $+1.08\%$, while ranking yields a $+1.12\%$ GMV/UU gain together with a $+2.90\%$ increase in advertising revenue.
Notably, in retrieval A/B testing, OnePiece covers nearly $70\%$ of impressions recalled by other strategies while delivering $2\times$ higher exclusive contribution than DLRM, demonstrating its ability to both subsume existing recall routes and provide substantial novel impressions and clicks.

Our main contributions can be summarized as follows:
\vspace{-0.3cm}
\begin{itemize}[leftmargin=0.5cm]
    \item To the best of our knowledge, this is the first work that explores and deploys context engineering and multi-step reasoning in industrial-scale ranking systems, achieving significant improvements over strong DLRM baselines in both retrieval and ranking tasks.
    \item We propose OnePiece, a unified framework that introduces both structured context engineering and block-wise latent reasoning into retrieval and ranking stages of cascaded pipelines, together with a progressive multi-task training strategy designed to effectively optimize multi-step reasoning.
    \item We conduct extensive offline and online evaluations, including large-scale A/B testing in Shopee’s main personalized search scenario. These experiments validate the effectiveness of each design choice, showcase favorable scaling and efficiency properties, and confirm the practicality of deploying OnePiece in real-world industrial environments.  
    
\end{itemize}

%% file: section/prelinminary.tex
\section{Preliminary}
In this section, we illustrate the problem formulation and typical retrieval-ranking cascade paradigm in industrial ranking systems. Without loss of generality, we focus on the personalized search setting where query-related information is present. In recommendation settings without explicit queries, the query-related component can be omitted.

\subsection{Problem Formulation}
A typical ranking system (\textit{e.g.}, a search or recommender system) aims to return an ordered list of items that the user is most likely to interact with next.
Formally, let $\mathcal{U}$, $\mathcal{V}$, and $\mathcal{Q}$ denote the sets of users, items, and queries (in search scenarios), respectively. For each user $u \in \mathcal{U}$, we define $\mathbf{u}$ as the complete user feature representation, encompassing both the user ID and user-side attributes (\textit{e.g.}, age, gender). Similarly, for each item $v \in \mathcal{V}$, we denote by $\mathbf{v}$ the complete item feature representation, comprising the item ID and item-side attributes (\textit{e.g.}, category, shop ID, price). In personalized search, each user query $q \in \mathcal{Q}$ has a feature representation $\mathbf{q}$ encoding the query ID and query-related Information (\textit{e.g.}, textual embeddings).
A user's historical behavior is modeled as a chronological sequence of interacted items: $S^u = (v_1^u, \ldots, v_t^u, \cdots v_{n_u}^u)$, where $v_t^u$ denotes the $t$-th item the user $u$ interacted with and $n_u$ is the number of interactions for user $u$.
Given a user's interaction history up to time $t$, denoted as $S_{1:t}^u = (v_1^u, \ldots, v_t^u)$, along with the current context features ($\mathbf{u}, \mathbf{v}, \mathbf{q}$), the personalized ranking system computes a relevance score for each candidate item $v \in \mathcal{V}$ through a learned scoring function:
\begin{equation}
\mathcal{F}_\theta(S^u_{1:t}, \mathbf{u}, \mathbf{v}, \mathbf{q}) \mapsto \mathbb{R},
\end{equation}
where $\theta$ represents the model parameters. The system subsequently produces a ranked list $\pi$ by sorting all candidate items in descending order of their scores. In pure recommendation scenarios without an explicit query, the user query $\mathbf{q}$ is excluded from the scoring function.

\subsection{Cascade Ranking Paradigm}
Cascade ranking has become the dominant paradigm in industrial large-scale ranking systems due to its ability to balance computational efficiency and ranking quality.
This paradigm organizes the decision process into multiple stages (\textit{e.g.}, Retrieval, Pre-ranking, Ranking) in a funnel-like manner: early stages employ lightweight models to rapidly filter out the majority of irrelevant items from massive candidate pools, while later stages apply increasingly sophisticated but computationally expensive models to progressively refined and smaller candidate sets, ultimately producing a high-quality ranked list.

For clarity, we consider a typical two-stage cascade ranking system consisting of:
\begin{itemize}[leftmargin=0.5cm]
    \item \textbf{Retrieval Stage.}
    The goal is to efficiently select a small set of promising candidates $\mathcal{V}^\prime$ from the full corpus $\mathcal{V}$.
    Industrial systems commonly adopt a dual-tower architecture: a \textit{user tower} transforms user context $(S^{u}_{1:t}, \mathbf{u}, \mathbf{q})$ into a dense representation $\mathbf{h}_u$, while an \textit{item tower} independently transforms each item $\mathbf{v}$ into $\mathbf{h}_v$.
    Retrieval is then performed via approximate nearest neighbor (ANN) search based on the similarity between user and item representations, typically computed as inner product or cosine similarity.
    \item \textbf{Ranking Stage.}
    The goal is to produce a finely-ordered list $\pi$ from retrieved candidates $\mathcal{V}^\prime$ by estimating precise preference scores.
    This stage typically adopts a single-tower architecture, where each candidate $v \in \mathcal{V}^\prime$ is jointly encoded with the full context $(S^{u}_{1:t}, \mathbf{u}, \mathbf{q})$ into a feature sequence. The sequence is processed by a Transformer-based backbone to model rich feature interactions, followed by an MLP head to estimate final scores.
\end{itemize}

In summary, the retrieval and ranking stages differ in both \textit{objectives} and \textit{inputs}:
retrieval focuses on coarse-grained recall with disjoint user/item encoders and the entire item corpus as the search space, while ranking performs fine-grained discrimination with joint encoding of the candidate items.

%% file: section/method.tex
\section{The Proposed OnePiece Framework}

In this section, we present \textbf{OnePiece}, a unified framework that brings LLM-style \textit{context engineering} and \textit{multi-step reasoning} into industrial cascaded ranking systems. We begin by introducing the overall design principles of OnePiece, then describe the structured context construction shared across the retrieval and ranking stages. We further detail the block-wise latent reasoning mechanism and the progressive multi-task optimization method.

\begin{figure}[t]  
    \centering  \includegraphics[width=1\columnwidth]{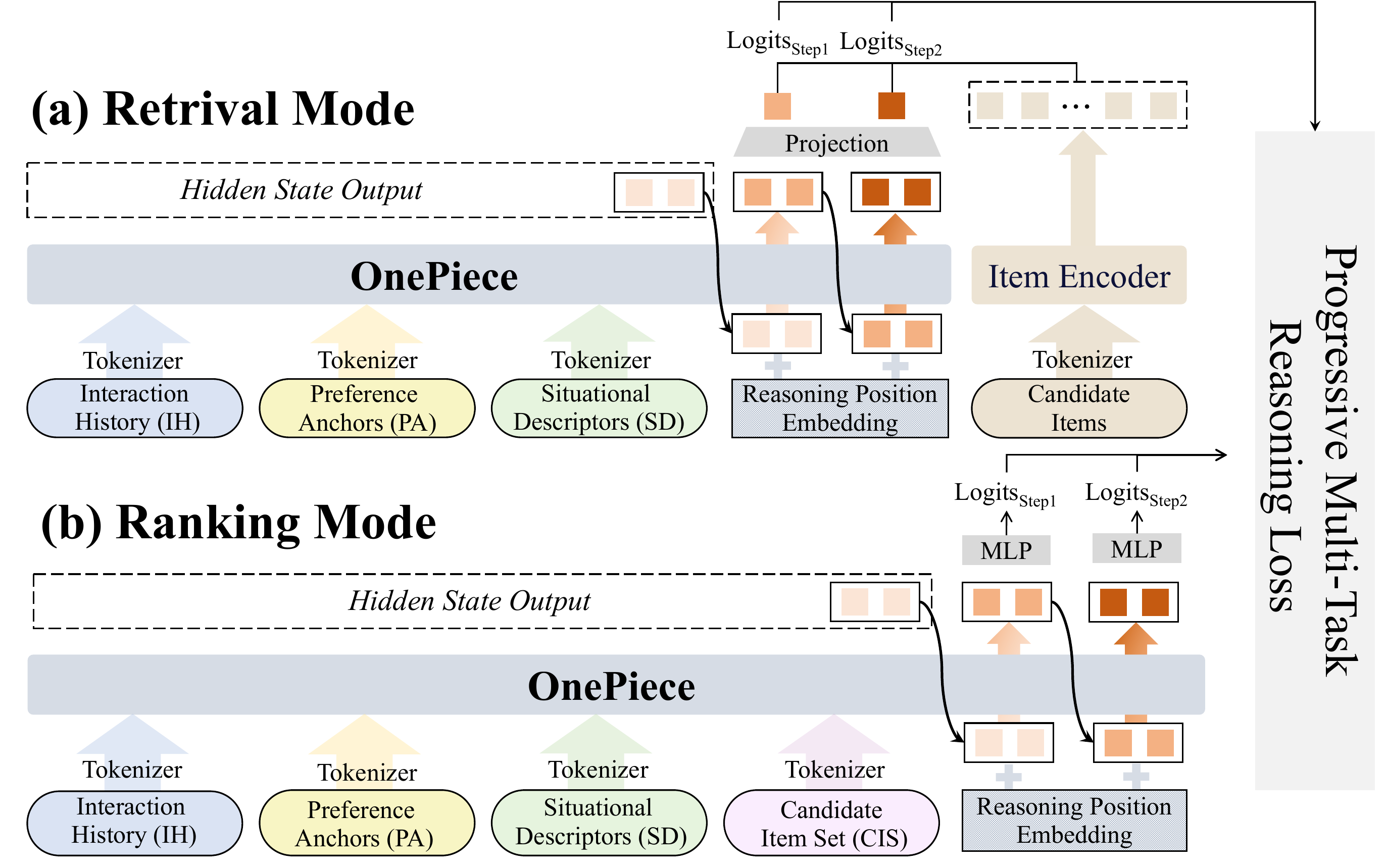}
   \caption{Overall architecture of the proposed \textbf{OnePiece} framework. Retrieval Mode~(a) and Ranking Mode~(b) both employ structured context engineering to construct unified input tokens, utilize block-wise latent reasoning to iteratively enhance representations across multiple reasoning steps, and are optimized through progressive multi-task training strategy.} 
    \label{fig:overall} 
\end{figure}

\subsection{Overview}
\textbf{OnePiece} is a unified framework for industrial cascaded ranking systems that combines \textit{structured context engineering}, \textit{block-wise latent reasoning}, and a \textit{progressive multi-task training strategy}:
\vspace{-0.3cm}
\begin{itemize}[leftmargin=0.5cm]
    \item \textbf{Context Engineering for Input Token Sequence.} A flexible LLM-style input construction that encodes heterogeneous signals into a unified token sequence, including (1) user interaction history, (2) preference anchors constructed according to expert experience, (3) situational descriptors capturing scenario-specific context, and (4) candidate item descriptors (used only in ranking).
    
    \item \textbf{Transformer-based Backbone with Block-wise Reasoning.} A simple and pure Transformer-based backbone augmented with latent reasoning blocks that iteratively refine intermediate representations, enabling step-by-step modeling of user preferences.
    
    \item \textbf{Progressive Multi-Task Training Strategy.} A staged optimization strategy that supervises different reasoning blocks with multi-level user feedback: earlier blocks are aligned with abundant weak signals (\textit{e.g.}, clicks), while later blocks are guided by stronger but sparser signals (\textit{e.g.}, purchases).  
\end{itemize}
\vspace{-0.2cm}

OnePiece unifies retrieval and ranking with a pure Transformer backbone, maintaining the efficiency required in industrial deployment while enhancing context-awareness and reasoning depth across cascade stages. In the following, we will illustrate the details of each component.

\begin{figure}[h]  
    \centering  \includegraphics[width=1\columnwidth]{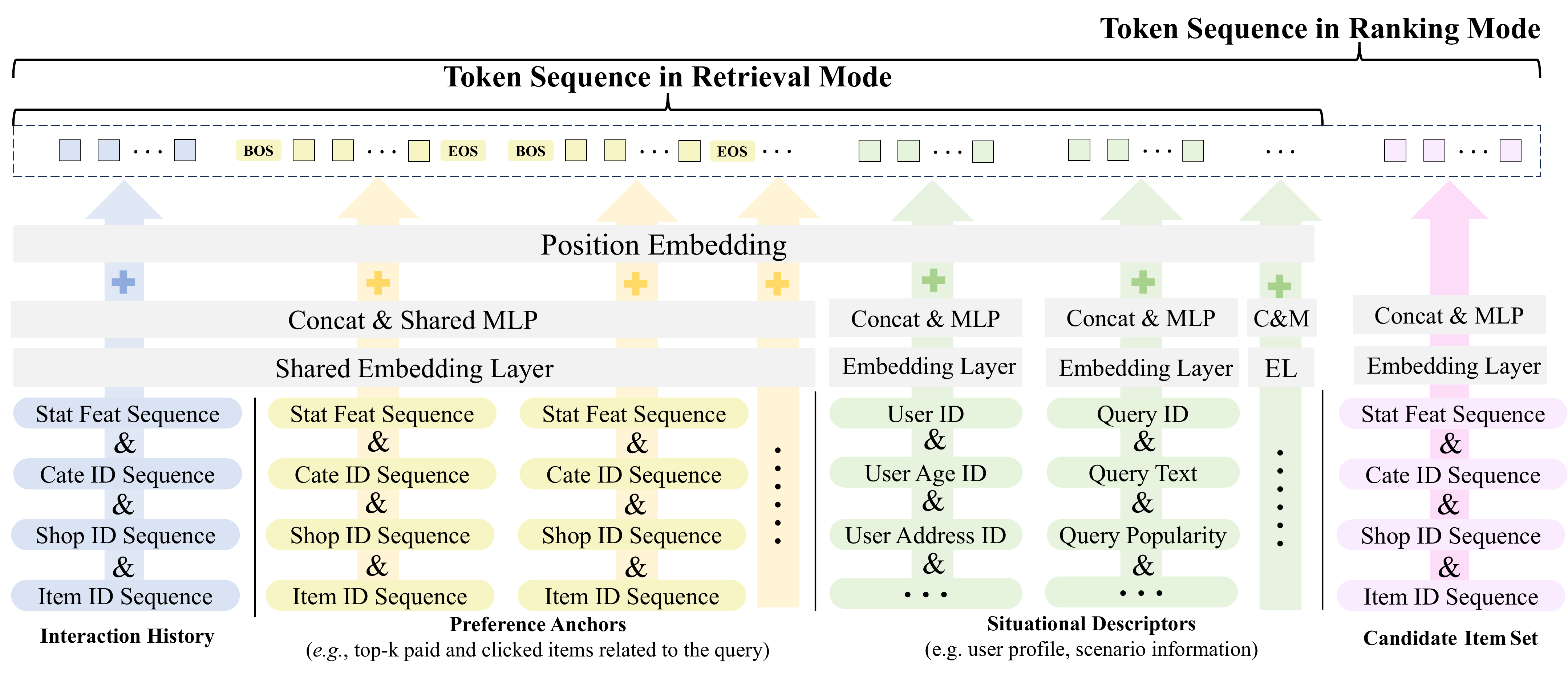}
    \caption{Context engineering and tokenizer design for input token sequences in \textbf{OnePiece}. 
    Both retrieval and ranking share the same construction of interaction history (IH), preference anchors (PA), and situational descriptors (SD). 
    The key difference is that ranking additionally incorporates candidate item set (CIS) tokens, enabling joint scoring within the single-tower architecture.}
    \label{fig:tokenizer} 
\end{figure}

\subsection{Context Engineering} \label{sec:context-engineering}

We transform all inputs into a unified \textit{token sequence} that can be directly processed by a transformer-based backbone.
To comprehensively capture the user intent and situational context information, we design four complementary token types:
\vspace{-0.2cm}
\begin{itemize}[leftmargin=*]
    \item \textbf{Interaction History (IH)} encodes the user's historical item interactions in chronological order, capturing temporal patterns and evolving preferences.
    \item \textbf{Preference Anchors (PA)} incorporate auxiliary item sequences constructed based on expert knowledge, providing informative cues to capture context-specific user intents beyond the interaction history.
    \item \textbf{Situational Descriptors (SD)} represent static user features and query-specific information, providing essential context for the current ranking task.
    \item \textbf{Candidate Item Set (CIS, ranking mode only)} contains complete feature representations of candidate items to enable fine-grained comparison and scoring.
\end{itemize}
This unified sequence representation enables the model to jointly reason over long-term behavioral patterns, domain reference knowledge, and contextual constraints within a unified framework.
\vspace{-0.2cm}

Following the preliminaries, we denote a user as $u$, a query (when present) as $q$, and an item as $v$. Their corresponding feature representations are denoted as $\mathbf{u}$, $\mathbf{q}$, and $\mathbf{v}$, respectively, where each representation contains the identifier and all associated content information.
We employ entity-specific embedding functions $\phi_{\text{user}}(\cdot)$, $\phi_{\text{query}}(\cdot)$, and $\phi_{\text{item}}(\cdot)$ to map each entity's information (including both categorical and continuous features) into a concatenated embedding vector. These embedding functions process different feature types and dimensions according to each entity's characteristics. To unify these embeddings into the $d$-dimensional hidden space of the backbone model, we use lightweight projection heads (\textit{e.g.}, MLPs). Specifically, we define $\text{Proj}_{\text{user}}$, $\text{Proj}_{\text{query}}$, and $\text{Proj}_{\text{cand}}$, where each projection maps from its corresponding input dimension to $\mathbb{R}^d$. The IH and PA components share $\text{Proj}_{\text{shared}}$, while user, query, and candidate tokens use their respective projection layers. 
Next, we detail each component of the input token sequence.

\textbf{Interaction History (IH).} 
Given the user’s chronological interaction sequence
$S^{u}=(\mathbf{v}^{u}_{1},\ldots,\mathbf{v}^{u}_{n_u})$, we embed each item descriptor using a shared projection layer:
\begin{equation}
\label{eq:ib-token}
\mathbf{z}^{\mathrm{IH}}_{t}
= \mathrm{Proj}_{\text{shared}}\big( \phi_{\mathrm{item}}(\mathbf{v}^{u}_{t}) \big) \in \mathbb{R}^{d}.
\end{equation}
We then incorporate temporal information by adding positional embeddings based on the interaction order:
\begin{equation}
\mathbf{h}^{\mathrm{IH}}_{t}=\mathbf{z}^{\mathrm{IH}}_{t}+\mathbf{p}_{t}^\mathrm{IH}, \quad 1\le t\le n_u,
\end{equation}
where $\mathbf{p}_t^{\text{IH}} \in \mathbb{R}^d$ denotes the learnable positional embedding for the $t$-th interaction in the sequence, and $\mathbf{h}_t^{\text{IH}}$ represents the final token embedding that integrates both content and temporal information.

\textbf{Preference Anchors (PA).} Preference anchors are constructed based on domain knowledge to guide the model’s reasoning process. The anchors serve as high-quality reference points that inject inductive biases into the sequence, thereby constraining the representation space and steering the model toward more plausible prediction directions. For instance, in personalized search scenarios, top-clicked items can be used as anchors to introduce collaborative filtering signals that reflect the aggregated behavior of many users, providing informative cues for capturing context-specific user intents.

Formally, let the current session of user $u$ provide $B$ anchor groups $\mathcal{A}^u = \{A_1^u, \ldots, A_B^u\}$ (\textit{e.g.}, top-$K$ purchased or clicked items related to the current user or query), where $A_b^u = (v_{b,1}^{PA}, \ldots, v_{b,m_b}^{PA})$ denotes the $b$-th group with $m_b$ items. For the $j$-th anchor in group $b$, we compute the token embedding as: 
\begin{equation}
\mathbf{z}^{\mathrm{PA}}_{b,j} = \mathrm{Proj}_{\text{shared}}\big(\phi_{\mathrm{item}}\big(\mathbf{v}^{\mathrm{PA}}_{b,j}\big)\big) \in \mathbb{R}^d, \quad \mathbf{h}_{b,j}^{\mathrm{PA}} = \mathbf{z}^{\mathrm{PA}}_{b,j} + \mathbf{p}_{j}^{\mathrm{PA}}.
\end{equation}
Here, $\mathbf{p}_j^{\mathrm{PA}} \in \mathbb{R}^d$ represents the positional embedding for the $j$-th position within a group.
To preserve group structure, we wrap each group with learnable boundary tokens $\mathbf{e}_{\mathrm{BOS}}, \mathbf{e}_{\mathrm{EOS}} \in \mathbb{R}^d$  and get the final token sequence for each anchor group:
\begin{equation}
(\mathbf{e}_{\mathrm{BOS}}, 
\ \mathbf{h}^{\mathrm{PA}}_{b,1}, \ldots, 
\ \mathbf{h}^{\mathrm{PA}}_{b,m_b}, 
\ \mathbf{e}_{\mathrm{EOS}}).
\end{equation}

\textbf{Situational Descriptors (SD).}
Situational context captures non-item information relevant to the ranking task. For the user $u$ with its associated features, we compute the embedding:
\begin{equation}
\mathbf{z}^\mathrm{U} = \mathrm{Proj}_{\text{user}}\big(\phi_{\mathrm{user}}(\mathbf{u})\big) \in \mathbb{R}^d, \quad \mathbf{h}^{\mathrm{U}} = \mathbf{z}^\mathrm{U} + \mathbf{p}_k^{\mathrm{U}},
\end{equation}
where $\mathbf{h}^{\mathrm{U}} \in \mathbb{R}^d$ represents the projected user embedding and $\mathbf{p}_k^{\mathrm{U}} \in \mathbb{R}^d$ denotes the positional embedding for the user token at position $k$ in the sequence.

Similarly, for the query $q$ (which can be omitted in recommendation scenarios), we obtain:
\begin{equation}
\mathbf{z}^{\mathrm{Q}} = \mathrm{Proj}_{\text{query}}\big(\phi_{\mathrm{query}}(\mathbf{q})\big) \in \mathbb{R}^d, \quad \mathbf{h}^{\mathrm{Q}} = \mathbf{z}^{\mathrm{Q}} + \mathbf{p}_k^{\mathrm{Q}},
\end{equation}
where $\mathbf{p}_k^{\mathrm{Q}} \in \mathbb{R}^d$ is the corresponding positional embedding for the query token.

\textbf{Candidate Item Set (CIS, ranking mode only).}
In the ranking stage, 
a key design choice in ranking is whether to process candidates in a \textit{pointwise} or \textit{setwise} manner.  
\begin{itemize} [leftmargin=0.5cm]
    \item In the \textit{pointwise} mode, each candidate item is concatenated with the user context and passed through the model independently. This design is computationally efficient and parallelizable across candidates, but it fails to directly compare between candidates, which is crucial for ranking. 
    \item In the \textit{full setwise} mode, the entire retrieved candidate set $\mathcal{V}'$ (often thousands of items) is encoded jointly. This allows the model to compare all candidates in a shared latent space, aligning better with the intrinsic nature of ranking. However, the computational cost is prohibitive, especially when combined with multi-step reasoning.  
\end{itemize}

To strike a balance between efficiency and expressiveness, we adopt a \textit{grouped setwise} strategy: the retrieved candidate set $\mathcal{V}'$ is randomly partitioned into smaller groups of size $C$ (\textit{e.g.}, 12). Each group is processed independently in the ranking mode, enabling intra-group interaction among candidates while keeping the per-group reasoning cost tractable. Importantly, since grouping is randomized across training and inference, the model learns to be robust to arbitrary candidate organizations and achieves stable scoring performance. In this way, grouped setwise ranking thus strikes a practical balance between pointwise ranking ($C=1$) and full-set ranking ($C=|\mathcal{V}'|$), making it suitable for large-scale real-world personalized ranking systems deployment.

Formally, given a candidate set $\mathcal{C}^u = \{\mathbf{v}^{\mathrm{CIS}}_{1}, \ldots, \mathbf{v}^{\mathrm{CIS}}_{C}\}$ containing $C$ items to be ranked, we embed each candidate item as:
\begin{equation}
\mathbf{z}^{\mathrm{CIS}}_{i} 
= \mathrm{Proj}_{\text{cand}}\big( \phi_{\mathrm{item}}(\mathbf{v}^{\mathrm{CIS}}_{i}) \big) \in \mathbb{R}^d.
\end{equation}
To avoid spurious correlations between position and relevance labels, we deliberately exclude positional embeddings for candidate tokens:
\begin{equation}
\mathbf{h}^{\mathrm{CIS}}_{i} = \mathbf{z}^{\mathrm{CIS}}_{i}, \quad 1 \le i \le C.
\end{equation}
This design choice ensures that the model evaluates each candidate purely based on its content rather than its position in the input sequence.

\textbf{Sequence Packing and Ordering.}  
Let $\oplus$ denote the concatenation of token subsequences.
The final input sequence to the backbone model is obtained by packing the Interaction Behaviors (IH), Preference Anchors (PA), Situational Descriptors (SD), and, in ranking mode only, the Candidate Item Set (CIS), following fixed ordering rules.

\begin{itemize} [leftmargin=0.5cm]
    \item \textit{Retrieval Mode.} The final input token sequence is constructed as:
    \begin{equation}
    \label{eq:retrieval-pack}
    \mathcal{I}^{u}_{\mathrm{retrieval}}
    = \underbrace{(\mathbf{h}^{\mathrm{IH}}_{1}, \ldots, \mathbf{h}^{\mathrm{IH}}_{n_u})}_{\text{chronological IH}}
    \ \oplus\ 
    \underbrace{\bigoplus_{b=1}^{B} \big( \mathbf{e}_{\mathrm{BOS}},\ \mathbf{h}^{\mathrm{PA}}_{b,1}, \ldots, \mathbf{h}^{\mathrm{PA}}_{b,m_b},\ \mathbf{e}_{\mathrm{EOS}} \big)}_{\text{PA groups ordered by business rule}}
    \ \oplus\ 
    \underbrace{(\mathbf{h}^{\mathrm{U}},\ \mathbf{h}^{\mathrm{Q}},\ \ldots)}_{\text{SD segment}}.
    \end{equation}
    Here: (1) IH tokens are ordered by ascending interaction timestamp; (2) Each PA group is wrapped by BOS/EOS boundary tokens, with groups ordered according to predefined business rules; (3) SD tokens have no temporal ordering and are placed in a segment with distinct positional indices.
    \item \textit{Ranking Mode.} We extend the retrieval-mode sequence by appending candidate item tokens:
    \begin{equation}
    \label{eq:ranking-pack}
    \mathcal{I}^{u}_{\mathrm{rank}}
    = \mathcal{I}^{u}_{\mathrm{retrieval}} 
    \ \oplus\ 
    \underbrace{(\mathbf{h}^{\mathrm{CIS}}_{1}, \ldots, \mathbf{h}^{\mathrm{CIS}}_{C})}_{\text{no positional embedding}}.
    \end{equation}
    CIS tokens are appended without positional encodings to prevent the model from learning spurious correlations between sequence positions and relevance labels during training.
\end{itemize}

\subsection{Backbone Architecture} \label{sec:backbone}

Our backbone consumes the packed token sequence from Sec.~\ref{sec:context-engineering} in a \textit{unified} manner for both retrieval and ranking.
Below, we first describe the sequence encoder and then the block-wise multi-step reasoning that exposes intermediate latents representations for multi-task progressive supervision learning discussed in Sec.~\ref{sec:PMT}.

\subsubsection{Transformer-Based Sequential Encoding}
Let $\mathcal{I}=[\mathbf{h}_1;\ldots;\mathbf{h}_N]$ denote the final input tokens constructed from Sec.~\ref{sec:context-engineering}, where $N$ is the length of the total input sequence.
We adopt an $L$-layer bi-directional Transformer~\citep{vaswani2017attention} with pre-normalization for the backbone architecture.
Formally, let $\mathbf{H}^{l}=[\mathbf{h}^{l}_1;\ldots;\mathbf{h}^{l}_N]\in\mathbb{R}^{N\times d}$ denote the hidden states at later $l$, with initial input $\mathbf{H}^{(0)}=\mathcal{I}$, the $l$-th layer ($1\le l\le L$) is
\begin{equation*}
\begin{aligned}
\mathbf{H}^{l}_{\mathrm{attn}} &= \mathbf{H}^{l-1} + \mathrm{MHSA}\!\left(\mathrm{LN}\!\left(\mathbf{H}^{l-1}\right)\right),\\
\mathbf{H}^{l} &= \mathbf{H}^{l}_{\mathrm{attn}} + \mathrm{FFN}\!\left(\mathrm{LN}\!\left(\mathbf{H}^{l}_{\mathrm{attn}}\right)\right),
\end{aligned}
\end{equation*}
where $\mathrm{MHSA}(\cdot)$ is multi-head self-attention with \textit{bi-directional} attention and $\mathrm{FFN}(\cdot)$ is a position-wise feed-forward network; $\mathrm{LN}(\cdot)$ denotes layer normalization (grouped LN can be used in practice).
We prefer bi-directional attention because the personalized ranking task is non-autoregressive at both training and serving modes, and allowing tokens to condition on the full context yields superior performance in practice. The final encoder output $\mathbf{H}^{L}=[\mathbf{h}^{L}_1;\ldots;\mathbf{h}^{L}_N]$ serves as the foundation for subsequent reasoning.

\subsubsection{Block-Wise Multi-Step Reasoning}
Inspired by recent work on reasoning-enhanced recommendation (\textit{e.g.}, ReaRec~\citep{tang2025think}), we also employ a multi-step refinement process. Different from prior approaches that recycle a single hidden state across iterations, we design a \textit{block-wise reasoning} mechanism, where a set of hidden states is iteratively transferred across steps. 
The motivation arises from the limitation of single-unit reasoning, which may overly compress signals due to the narrow capacity of its transmission medium, thereby losing important information. In contrast, block-wise reasoning offers adjustable reasoning bandwidth, providing greater flexibility and achieving a better balance between information compression and retention. Moreover, by grouping tokens into reasoning blocks, the model is encouraged to assign specialized roles to different tokens, leading to more structured and effective refinement of representations.

\begin{wrapfigure}{r}{0.5\linewidth}  
    \centering
    \vspace{-2em}
    \includegraphics[width=\linewidth]{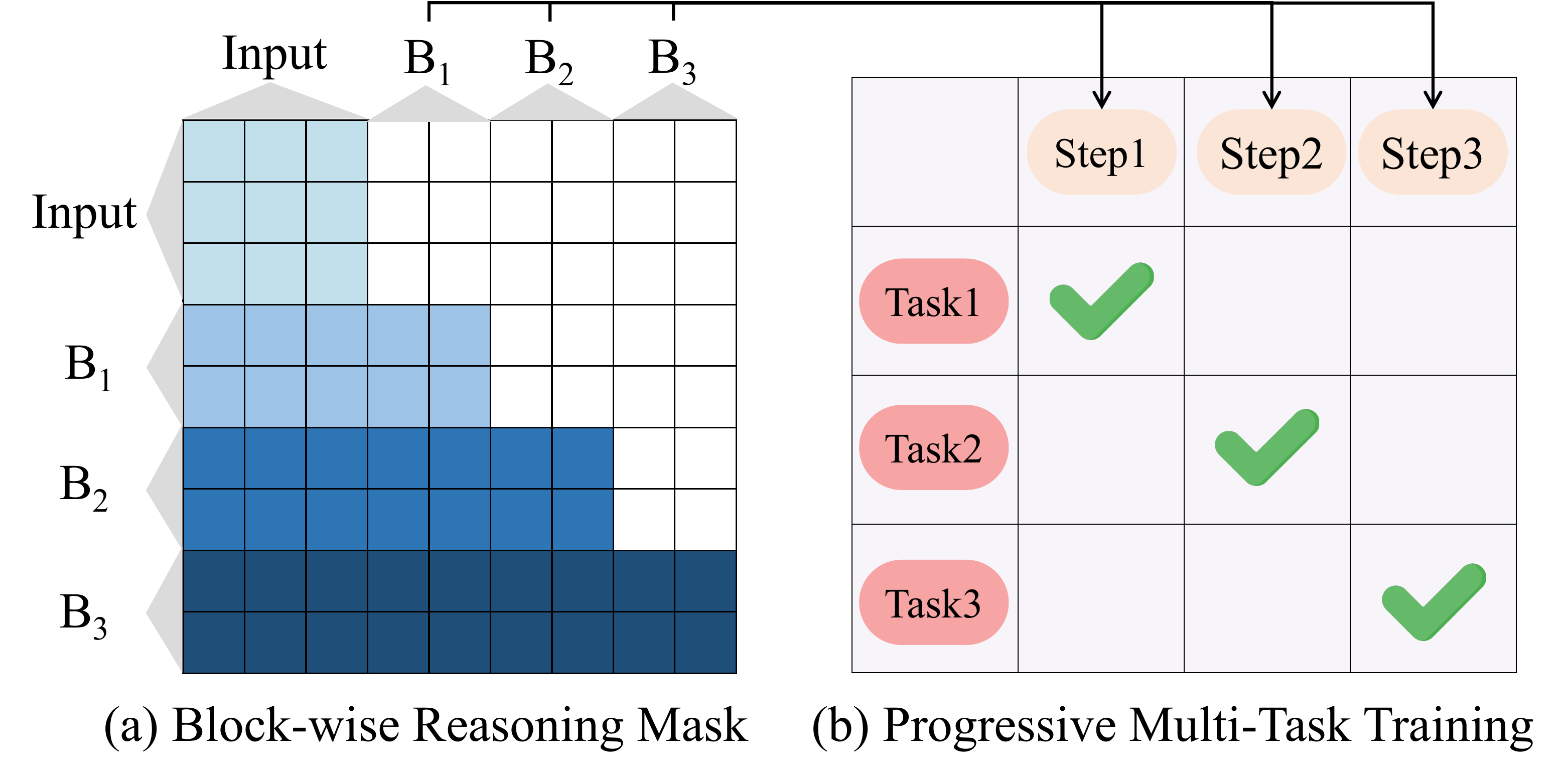}  
    \caption{Block-wise reasoning mask and progressive multi-task training. \textbf{(a)} Causal attention mask enables reasoning blocks to attend to input and previous blocks. \textbf{(b)} Progressive training assigns tasks of increasing complexity to successive reasoning steps to provide effective process supervision.}
    \label{fig:attention_mask}
\end{wrapfigure}

Formally, let $M$ denotes the block size which is task-dependent, and let $\mathbf{B}_k \in \mathbb{R}^{M\times d}$ denote the $k$-th ($k \in \mathbb{Z}$) reasoning block. 
The initial block $\mathbf{B}_0$ is constructed from the final encoder output $\mathbf{H}^{L}$ as
\[
\mathbf{B}_0 = \mathbf{H}^{L}[N-M+1:N] \in \mathbb{R}^{M\times d}.
\]

For step $k\ge 1$, we first extract the block from the previous reasoning step output:
\[
\mathbf{B}_k = \mathbf{H}_{k-1}^{L}[N+(k-2)M+1 : N+(k-1)M] \in \mathbb{R}^{M\times d}.
\]

To distinguish different reasoning steps, following the design of ReaRec~\cite{tang2025think}, we introduce Reasoning Position Embeddings (RPE). Let $\mathbf{E}_{\text{RPE}} \in \mathbb{R}^{K \times d}$ be a learnable embedding matrix where $K$ is the maximum number of reasoning steps. We define:
\begin{align*}
\tilde{\mathbf{B}}_0 &= \mathbf{B}_0, \\
\tilde{\mathbf{B}}_k &= \mathbf{B}_k + \mathbf{1}_M \otimes \mathbf{E}_{\text{RPE}}[k,:] \quad \text{for } k \geq 1,
\end{align*}
where $\mathbf{1}_M \in \mathbb{R}^M$ is a vector of ones and $\otimes$ denotes the outer product.
At each step $k$, we concatenate the base sequence $\mathcal{I}$ with all previous enhanced blocks $\tilde{\mathbf{B}}_{< k}$ and the current block $\tilde{\mathbf{B}}_k$, then pass them through the Transformer backbone with a block-wise causal mask:
\begin{equation}
\label{eq:reasoning-step}
[\mathcal{I};\tilde{\mathbf{B}}_{< k};\tilde{\mathbf{B}}_k] 
\;\xrightarrow{\;\;\mathcal{F}_{\theta}(\cdot;\mathcal{M}_k)\;\;}\; 
\mathbf{H}^L_k \in \mathbb{R}^{(N+kM)\times d},
\end{equation}
where $\mathcal{F}_{\theta}$ is the Transformer update and $\mathcal{M}_k$ is the mask mechanism that enforces information flow constraints.
As shown in Figure~\ref{fig:attention_mask}(a), we adopt a causal block-wise mask $\mathcal{M}_k$, which enforces that  
(1) block tokens $\tilde{\mathbf{B}}_k$ can attend to all base tokens $\mathcal{I}$ and all historical blocks $\tilde{\mathbf{B}}_{<k}$;  
(2) tokens of current reasoning block cannot attend to future reasoning block tokens.  
Iteratively applying this procedure yields progressively refined reasoning states $\tilde{\mathbf{B}}_1, \tilde{\mathbf{B}}_2, \ldots, \tilde{\mathbf{B}}_K$.

The above formulation is generic, and we adopt task-specific strategies for block size $M$ in industrial cascaded systems:
\begin{itemize}[leftmargin=0.6cm]
    \item \textit{Retrieval Mode.}  
    Here the block size $M$ is set equal to the length of the situation descriptor (SD), which provides enough capacity for compact yet expressive refinement. Since retrieval serves as a coarse filtering stage, the objective is to strengthen and balance scenario-specific goals. Taking personalized search as an example, we designate the user ($\mathbf{h}_U$) and query ($\mathbf{h}_Q$) tokens as aggregation blocks: $\mathbf{h}_U$ encodes personalization signals while $\mathbf{h}_Q$ captures relevance, and iterative reasoning over these tokens allows the model to jointly reinforce both dimensions during matching.
    \item \textit{Ranking Mode.} 
    Here the objective is to distinguish subtle differences within a constructed candidate set, and the candidate item set (CIS) is fully visible to the model. We therefore set $M=C$, where each block corresponds to all candidate item tokens. item tokens. The final block $\tilde{\mathbf{B}}_K$ thus contains the refined representations used for ranking. To further prevent overfitting to candidate order and encourage robust set-wise reasoning, we apply randomized candidate grouping during training, which ensures the model learns reasoning abilities invariant to specific candidate arrangements within each group.
\end{itemize}

\subsection{Progressive Multi-Task Training}\label{sec:PMT}
Building upon the block-wise multi-step reasoning framework, we obtain a sequence of intermediate block representations $\{\mathbf{B}_k\}_{k=1}^{K}$, where $\mathbf{B}_k \in \mathbb{R}^{M \times d}$ denotes the $k$-th reasoning block containing $M$ token states of dimension $d$. To provide effective process supervisory signals for the multi-step reasoning trajectory, we introduce a progressive multi-task training paradigm that implements curriculum learning~\citep{bengio2009curriculum} through gradually increasing task complexity.

We define $K$ learning objectives $\mathcal{T} = \{\tau_1, \tau_2, \ldots, \tau_{K}\}$ arranged in a progressive curriculum from fundamental to advanced tasks. For example, in e-commerce domains, this follows the natural user engagement progression: exposure $\to$ click $\to$ purchase, where each successive task requires increasingly sophisticated understanding of user preferences and behavioral patterns. Each reasoning step $k$ is assigned to optimize a single task $\tau_k$, creating a step-by-step learning trajectory where the model first masters basic recognition capabilities before advancing to complex preference modeling. This progressive assignment provides rich process supervision at each reasoning stage, enabling more effective learning of intermediate representations and ensuring that each reasoning block develops specialized capabilities while building upon the foundation established by previous steps.

\subsubsection{Retrieval Mode}
In the retrieval stage, we extract user representations from the sequence of reasoning blocks and optimize them using a combination of calibrated probability estimation and bidirectional contrastive learning objectives.

Specifically, given the sequence of reasoning blocks $\{\mathbf{B}_k\}_{k=1}^{K}$ where $\mathbf{B}_k \in \mathbb{R}^{M \times d}$, we extract a step-specific user representation from each block via layer normalization followed by mean pooling:
\[
    \mathbf{r}_k = \mathrm{Mean}(\mathrm{LN}(\mathbf{B}_k)) \in \mathbb{R}^d, \quad k \in \{1,2,\ldots,K\}.
\]
For each training instance, we construct a candidate pool $\Omega$ containing items with diverse behavioral labels across the progressive task sequence. Let $\mathbf{z}_v \in \mathbb{R}^d$ denote the item embedding for candidate $v \in \Omega$. For the task $\tau_k$ assigned to step $k$ and candidate $v \in \Omega_{\tau_k}$, let $y^k_v \in \{0,1\}$ denote the corresponding behavioral label. We partition the candidate pool into positive and negative sets:
\begin{align*}
    \Omega_{\tau_k}^+ = \{v \in \Omega_{\tau_k} : y^k_v = 1\}, \quad
    \Omega_{\tau_k}^- = \{v \in \Omega_{\tau_k} : y^k_v = 0\}.
\end{align*}

We employ two complementary learning objectives that address different granularities of representation optimization:

\noindent
\textbf{(i) Binary Cross-Entropy Loss (BCE)} operates at the point-wise level, providing calibrated probability estimates for individual user-item pairs:
\begin{equation}
    \mathcal{L}^{\mathrm{BCE}}_k =
    \sum_{v \in \Omega_{\tau_k}^+} -\log \sigma(\langle \mathbf{r}_k, \mathbf{z}_v \rangle) +
    \sum_{v \in \Omega_{\tau_k}^-} -\log \big(1-\sigma(\langle \mathbf{r}_k, \mathbf{z}_v \rangle)\big),
\end{equation}
where $\sigma(\cdot)$ is the sigmoid function.

\noindent
\textbf{(ii) Bidirectional Contrastive Learning (BCL)} operates at the batch level, enabling global contrastive reasoning across the in-batch samples. Drawing inspiration from CLIP~\citep{radford2021learning}, we propose BCL optimization objective that consists of two symmetric components:

\textbf{User-to-Item (U2I) Contrastive Learning} enables each user representation $\mathbf{r}_k$ to holistically distinguish positive items from negative candidates, which is formulated as:
\begin{equation}
    \mathcal{L}^{\mathrm{U2I}}_k = \sum_{v \in \Omega_{\tau_k}^+} -\log 
    \frac{\exp(\langle \mathbf{r}_k, \mathbf{z}_v \rangle / \eta)}
    {\sum_{v^+ \in \Omega_{\tau_k}^+} \exp(\langle \mathbf{r}_k, \mathbf{z}_{v^+} \rangle / \eta) +
    \sum_{v^- \in \Omega_{\tau_k}^-} \exp(\langle \mathbf{r}_k, \mathbf{z}_{v^-} \rangle / \eta)},
\end{equation}
where $\eta > 0$ is the temperature parameter.

\textbf{Item-to-User (I2U) Contrastive Learning} enables each positive item to identify its corresponding user from all in-batch users. Let $\mathcal{R}_k = \{\mathbf{r}_k^{(i)}\}_{i=1}^B$ denote the set of step-$k$ user representations within the current training batch, where $B$ is the batch size. Then, the I2U objective is:
\begin{equation}
    \mathcal{L}^{\mathrm{I2U}}_k = \sum_{v \in \Omega_{\tau_k}^+} -\log 
    \frac{\exp(\langle \mathbf{r}_k, \mathbf{z}_v \rangle / \eta)}
    {\sum_{\mathbf{r}' \in \mathcal{R}_k} \exp(\langle \mathbf{r}', \mathbf{z}_v \rangle / \eta)}.
\end{equation}

The complete BCL objective for step $k$ combines both symmetric components:
\begin{equation}
    \mathcal{L}^{\mathrm{BCL}}_k = 
    \mathcal{L}^{\mathrm{U2I}}_k + \mathcal{L}^{\mathrm{I2U}}_k.
\end{equation}
And the overall retrieval loss aggregates the objectives across all reasoning steps, where each step $k$ optimizes for its assigned task $\tau_k$:
\begin{equation}
    \mathcal{L}^{\mathrm{retrieval}} = 
    \sum_{k=1}^{K} \big( \mathcal{L}^{\mathrm{BCE}}_k + \mathcal{L}^{\mathrm{BCL}}_k \big).
\end{equation}
This progressive training strategy provides comprehensive process supervision, where each reasoning step receives dedicated supervisory signals tailored to its assigned task complexity, achieving more effective learning of multi-step reasoning capabilities.

\subsubsection{Ranking Mode}
In the ranking stage, the block size equals the candidate group size ($M = C$). Each reasoning block $\mathbf{B}_k \in \mathbb{R}^{C \times d}$ contains $C$ hidden states $\{\mathbf{h}_{i,k}\}_{i=1}^C$, where $\mathbf{h}_{i,k} \in \mathbb{R}^d$ represents the hidden state for the $i$-th candidate at reasoning step $k$.
For the task $\tau_k$ assigned to reasoning step $k$, we compute candidate-wise logits through a task-specific scoring network:
\[
s_{i,k} = \mathrm{MLP}_{\tau_k}(\mathbf{h}_{i,k}) \in \mathbb{R}, \quad i \in \{1,\ldots,C\},
\]
where $\mathrm{MLP}_{\tau_k}$ denotes the Multi-Layer Perceptron (MLP) for task $\tau_k$.

We employ two complementary learning objectives:

\noindent
\textbf{(i) Binary Cross-Entropy Loss (BCE)} provides point-wise probability calibration for individual candidates:
\begin{equation}
\mathcal{L}^{\mathrm{BCE}}_k = \sum_{i=1}^{C} \left[ -y^{\tau_k}_i \log \sigma(s_{i,k}) - (1 - y^{\tau_k}_i) \log(1 - \sigma(s_{i,k})) \right],
\end{equation}
where $y^{\tau_k}_i \in \{0,1\}$ denotes the behavioral label of candidate $i$ for task $\tau_k$.

\noindent
\textbf{(ii) Set Contrastive Learning (SCL)} operates at the set-wise level, enabling each positive candidate to distinguish itself from negative candidates within the set:
\begin{equation}
\mathcal{L}^{\mathrm{SCL}}_k = \sum_{i: y^{\tau_k}_i = 1} -\log 
\frac{\exp(s_{i,k} / \eta)}
{\sum_{j=1}^{C} \exp(s_{j,k} / \eta)},
\end{equation}
where the summation is over all positive candidates, and each positive candidate competes against all candidates in the group for ranking position.

Then, the overall ranking loss combines both objectives across all reasoning steps:
\begin{equation}
\mathcal{L}^{\mathrm{ranking}} = \sum_{k=1}^{K} \left( \mathcal{L}^{\mathrm{BCE}}_k + \mathcal{L}^{\mathrm{SCL}}_k \right).
\end{equation}
This curriculum-based multi-task training provides dense supervision signals throughout the reasoning process, enabling the model to learn more effective step-by-step reasoning patterns while ensuring each reasoning block develops appropriate capabilities for its designated task complexity level.

\subsection{Time Complexity Analysis}  
In this section, we analyze the complexity of our block-wise reasoning framework under retrieval and ranking modes.  

For the backbone encoder (without reasoning), the time complexity of each Transformer layer is  
\[
O(N^2d + Nd^2),
\]  
where $N$ is the base sequence length, $d$ is the hidden dimension, and $L$ is the number of layers. Thus the total cost is $O(L(N^2d + Nd^2))$.  

For the reasoning phase, we employ the \textit{KV Caching} technique to reuse historical key–value pairs, so that each new reasoning step only incurs attention between the $M$ new block tokens and the cached tokens. Specifically, at reasoning step $k$, we need to: 
(i) Compute Q, K, V for $M$ new block tokens: $O(Md^2)$; 
(ii) Calculate attention between $M$ new tokens and all $N + (k-1)M$ cached tokens: $O(M(N + kM)d)$; 
(iii) Apply output projection: $O(Md^2)$. 
Therefore, the complexity per layer at step $k$ is:
\[
    O\big(M(N + kM)d + Md^2\big).
\] 
Aggregating over $K$ reasoning steps and $L$ layers, the total additional reasoning cost is 
\[
O\big(LKM(Nd + MKd + d^2)\big).
\]  

%% file: section/experiment.tex
\section{Offline Experiments}
To validate the effectiveness of OnePiece, we first conduct extensive offline experiments based on 30-day log from Shopee, an e-commercial platform serving billion-scale users across Southeast Asia and Latin America. 
The full dataset statistics are summarized in Table~\ref{tab:shopee_dataset_stats}, and the detailed construction process of training samples is described in Appendix~\ref{app:offline_dataset_cons}.
Furthermore, we perform comprehensive ablation studies and other detailed analyses to reveal the advantages of OnePiece.

\begin{table}[htbp]
\centering
\caption{Dataset Statistics of Shopee E-commerce Platform from June 11 to July 10, 2025. The dataset includes multi-behavior user interaction. Abbreviations: M = Million (10\textsuperscript{6}), B = Billion (10\textsuperscript{9}).}
\label{tab:shopee_dataset_stats}
\begin{tabular}{@{}cccccccc@{}}
\toprule
\textbf{\#User} & \textbf{\#Item} & \textbf{\#Query} & \textbf{\#Impression} & \textbf{\#Click} & \textbf{\#Add-to-Cart} & \textbf{\#Order} \\
\midrule
10M & 93M & 12M & 0.24B & 60M & 12M & 6M \\
\bottomrule
\end{tabular}
\end{table}

\subsection{Experimental Settings}

\subsubsection{Evaluation Protocol}
We adopt a streaming-style training and evaluation protocol to closely mimic real-world deployment. 
For training, we employ an incremental scheme where the model is trained day by day, with samples within each day randomly shuffled to ensure robustness. 
For offline evaluation, we follow a strict chronological order: the checkpoint trained on day-$t$ is evaluated on the unseen data from day-$(t+1)$. 
Unless otherwise specified, all experiments are conducted on the same $30$ consecutive days of data to ensure comparability across methods.

\subsubsection{Evaluation Metrics} For the retrieval stage, we are primarily concerned with how many clicked items are successfully recalled. 
Thus, we report Recall@100 and Recall@500, denoted as R@100 and R@500. 
For the ranking stage, we evaluate the model under three types of user feedback: click (C-), add-to-cart (A-), and order (O-). 
For each type, we report both AUC and GAUC, denoted as C-AUC/C-GAUC, A-AUC/A-GAUC, and O-AUC/O-GAUC, respectively.

\subsubsection{Baseline Methods}
We compare OnePiece against several representative baselines:  

\textbf{DLRM} (Production baseline in Shopee): Shopee’s DLRM baseline is a well-optimized hybrid model including multiple SOTA components. For \textit{retrieval mode}, it adopts a two-tower design inspired by DSSM~\citep{huang2013learning}, where the query and user history are encoded separately. To enhance relevance modeling, it employs a DIN-like~\citep{zhou2018deep} structure and zero-attention~\citep{ai2019zero}. A lightweight text CNN is used to extract keyword features, while DCNv2~\citep{wang2021dcn} captures high-order cross features. Sequential features are aggregated by mean pooling, concatenated with other features, and fused via an MLP. For \textit{ranking mode}, the DLRM model is different because we can allow incorporating the candidate item into a single tower with user features. The backbone is ResFlow~\citep{fu2024residual}, combined with DIN-like target attention and cross-attention across sequential behaviors. DCNv2 is again used for higher-order interactions, followed by MLP fusion. A SENet~\citep{hu2017squeeze} module further supports adaptive feature selection for different tasks.

\textbf{HSTU}~\citep{zhai2024actions}:  
HSTU is a representative generative recommendation framework proposed by Meta. For a fair comparison, we align its parameter size with OnePiece and adopt the same side-information fusion strategy. Since the original HSTU only considers Interaction History (IH) and Situational Descriptors (SD), we additionally evaluate a variant HSTU+PA where Preference Anchors (PA) are introduced into its input sequence, consistent with OnePiece’s context engineering.

\textbf{ReaRec}~\citep{tang2025think}:  
ReaRec is a representative reasoning-enhanced recommendation model that formulates user representation modeling as multi-step reasoning over item sequences. The vanilla ReaRec architecture only supports retrieval tasks with the user iteration history sequence as input. To ensure experimental consistency, we adapt its backbone and feature inputs to match OnePiece. For ranking, we adapt ReaRec by introducing the candidate item into the input sequence and applying a target-aware attention mask (following the design in HSTU), where sequence tokens can attend to the candidate but candidates remain mutually invisible. Similar to HSTU, we also evaluate a ReaRec+PA variant that augments IH and SD with Preference Anchors, providing a stronger context prior for reasoning.

\begin{table}[t]
\centering
\caption{Performance comparison of different models on both retrieval and ranking tasks with 30 days of training data. 
For retrieval, we report Recall@100 (R@100) and Recall@500 (R@500) evaluated on clicked samples. 
For ranking, we report AUC and GAUC under three feedback types: click (C-), add-to-cart (A-), and order (O-). 
The best results are highlighted in \textbf{bold}.}
\label{tab:exp_offline_main_results}
\resizebox{0.95 \linewidth}{!}{
\begin{tabular}{lcccccccc}
\toprule
\multirow{2}{*}{Model} & \multicolumn{2}{c}{Retrieval Mode} & \multicolumn{6}{c}{Ranking Mode} \\
\cmidrule(lr){2-3} \cmidrule(lr){4-9}
& R@100 & R@500 & C-AUC & C-GAUC & A-AUC & A-GAUC & O-AUC & O-GAUC \\
\midrule
DLRM        & 0.458 & 0.679 & 0.856 & 0.851 & 0.893 & 0.843 & 0.931 & 0.854 \\
HSTU        & 0.443 & 0.658 & 0.833 & 0.829 & 0.878 & 0.827 & 0.913 & 0.839 \\
HSTU+PA     & 0.472 & 0.680 & 0.855 & 0.852 & 0.901 & 0.848 & 0.926 & 0.849 \\
ReaRec      & 0.452 & 0.674 & 0.843 & 0.838 & 0.882 & 0.834 & 0.919 & 0.843 \\
ReaRec+PA   & 0.485 & 0.701 & 0.862 & 0.863 & 0.908 & 0.851 & 0.927 & 0.851 \\
\rowcolor{orange!30}
\textbf{OnePiece} 
            & \textbf{0.517} & \textbf{0.731} & \textbf{0.911} & \textbf{0.909} & \textbf{0.952} & \textbf{0.897} & \textbf{0.963} & \textbf{0.886} \\
\bottomrule
\end{tabular}
}
\end{table}

\subsection{Overall Performance}

Table~\ref{tab:exp_offline_main_results} summarizes the performance of different models on both retrieval and ranking tasks. Several observations can be drawn as follows:   

First, \textbf{DLRM remains a very strong baseline}. Compared with original HSTU and ReaRec, DLRM achieves better performance on most evaluated metrics because it fully exploits rich feature interactions, target attention, and multiple sequential features. In contrast, the original HSTU and ReaRec underperform DLRM as they only rely on interaction behavior sequences as input.  

Second, \textbf{the preference anchor (PA) mechanism brings consistent benefits independent of the backbone}. Both HSTU+PA and ReaRec+PA outperform their vanilla counterparts, confirming that enriching user history with auxiliary preference signals provides complementary information. Between the two, ReaRec demonstrates higher robustness due to its Transformer backbone with bidirectional attention and reasoning capability, which enables it to better leverage the anchor information than HSTU.  

Finally, \textbf{OnePiece achieves the best overall results}. Compared with the strongest ReaRec+PA baseline, OnePiece further improves Recall@100 from $0.485$ to $0.517$ and C-AUC from $0.862$ to $0.911$. These consistent gains can be attributed to its novel block-wise latent reasoning and the progressive multi-task training strategy, which enable finer-grained preference refinement through multi-step reasoning. Together, these advances validate the design of OnePiece as a more powerful reasoning-enhanced recommendation framework capable of unifying retrieval and ranking.

\subsection{Ablation Study}
In this section, we provide detailed ablation studies to evaluate the contribution of each design in OnePiece through a step-by-step analysis.

\subsubsection{Context Engineering Ablation}

\begin{table}[t]
\centering
\caption{Ablation studies on context engineering of OnePiece. PA($L$) denotes appending a preference-anchor sequence with maximum length $L$. SD denotes situational descriptors.}
\label{tab:ablation_context}
\resizebox{\linewidth}{!}{
\begin{tabular}{l l cc cccccc}
\toprule
\multirow{2}{*}{Version} & \multirow{2}{*}{Model} & \multicolumn{2}{c}{Retrieval} & \multicolumn{6}{c}{Ranking} \\
\cmidrule(lr){3-4} \cmidrule(lr){5-10}
& & R@100 & R@500 & C-AUC & C-GAUC & A-AUC & A-GAUC & O-AUC & O-GAUC \\
\midrule
\rowcolor{orange!5}
V1 & IH(ID)              & 0.407 & 0.646 & 0.802 & 0.802 & 0.860 & 0.819 & 0.908 & 0.835 \\
\rowcolor{orange!5}
V2 & IH(ID+Side Info)    & 0.428 & 0.657 & 0.846 & 0.844 & 0.871 & 0.839 & 0.918 & 0.845 \\
\rowcolor{orange!15}
V3 & V2+PA(10)           & 0.459 & 0.677 & 0.879 & 0.876 & 0.923 & 0.863 & 0.940 & 0.861 \\
\rowcolor{orange!15}
V4 & V2+PA(20)           & 0.467 & 0.686 & 0.885 & 0.886 & 0.929 & 0.869 & 0.946 & 0.866 \\
\rowcolor{orange!15}
V5 & V2+PA(30)           & 0.475 & 0.689 & 0.892 & 0.890 & 0.936 & 0.874 & 0.949 & 0.871 \\
\rowcolor{orange!15}
V6 & V2+PA(60)           & 0.491 & 0.707 & 0.901 & 0.900 & 0.945 & 0.886 & 0.956 & 0.880 \\
\rowcolor{orange!15}
V7 & V2+PA(90)           & 0.504 & 0.719 & 0.908 & 0.905 & 0.951 & 0.896 & 0.962 & 0.885 \\
\rowcolor{orange!30}
V8 & V7+SD               & \textbf{0.517} & \textbf{0.731} & \textbf{0.911} & \textbf{0.909} & \textbf{0.952} & \textbf{0.897} & \textbf{0.963} & \textbf{0.886} \\
\bottomrule
\end{tabular}
}
\end{table}

To validate the effectiveness of our context engineering design, we conduct a step-by-step ablation study, progressively adding Interaction History (IH), Preference Anchors (PA), and Situational Descriptors (SD) into the model input. The experiment results for both retrieval and ranking are summarized in Table~\ref{tab:ablation_context}.

We begin with a minimal baseline (V1), where only user interaction sequences composed of raw item IDs are utilized. This configuration, implemented with a two-layer Transformer and bidirectional attention, yields the lowest performance across both retrieval and ranking tasks. Furthermore, introducing side information for each item (V2) leads to a clear improvement, highlighting the importance of more item features beyond raw IDs.

Building on this, we incorporate Preference Anchors (PA) by appending users' query-associated item sequences comprising top-$k$ purchases, top-$k$ clicks, and top-$k$ impressions. From V3 to V7, we gradually increase the sequence length from 10 to 90, progressively extending the combined behavioral sequence. The results demonstrate a clear \textbf{scaling effect of PA}: extending the auxiliary item sequence enriches query-specific context, enabling the model to capture more fine-grained user intent and leading to steadily higher performance. In particular, PA provides query-dependent signals that are absent in plain IH, allowing the model to differentiate user preferences under different queries rather than relying solely on global interaction history or generic query descriptors.  

Finally, incorporating Situational Descriptors (V8) yields the best overall results, especially on retrieval. Compared with V7, Recall@100 improves from $0.504$ to $0.517$, as the additional user- and query-side descriptors provide stronger contextual grounding that helps retrieve a broader set of relevant items. By contrast, the ranking gains are marginal, since IH already supplies rich personalization signals and PA captures detailed query-specific preferences. Given that ranking primarily focuses on fine-grained comparisons among candidate items where query relevance is high, SD serves as a weaker anchor in this stage.

Overall, these ablation studies demonstrate that \textbf{structured context engineering is highly effective}: IH captures long-term personalization, PA provides scalable and query-specific anchors, and SD contributes stable contextual grounding, particularly benefiting retrieval. Together, they enrich user–query representations in complementary ways and enable OnePiece to achieve consistent improvements across retrieval and ranking.  

\subsubsection{Training Strategy Ablation}

\begin{table}[t]
\centering
\caption{Effect of different training strategies on retrieval performance. Here, multi-task refers to jointly optimizing impression and click prediction losses.}
\label{tab:ablation_training_retrieval}
\resizebox{0.7\linewidth}{!}{
\begin{tabular}{l l cc}
\toprule
Version & Training Strategy & R@100 & R@500 \\
\midrule
\rowcolor{orange!5}
V1 & Causal Mask                                        & 0.464 & 0.671 \\
\rowcolor{orange!5}
V2 & Bi-Directional               & 0.470 & 0.676 \\
\rowcolor{orange!15}
V3 & V2 + 1-Step Reasoning, Click Task on Last Step        & 0.490 & 0.708 \\
\rowcolor{orange!15}
V4 & V2 + 1-Step Reasoning, Multi-Task on Last Step      & 0.495 & 0.714 \\
\rowcolor{orange!15}
V5 & V2 + 2-Step Reasoning, Multi-Task on Last Step      & 0.510 & 0.726 \\
\rowcolor{orange!30}
V6 & V2 + 2-Step Reasoning, Progressive Multi-Task                 & \textbf{0.517} & \textbf{0.731} \\
\bottomrule
\end{tabular}
}
\end{table}

\begin{table}[t]
\centering
\caption{Effect of different training strategies on ranking performance. Here, multi-task refers to jointly optimizing impression, click, and order prediction losses.}
\label{tab:ablation_training_ranking}
\resizebox{\linewidth}{!}{
\begin{tabular}{l l cccccc}
\toprule
Version & Training Strategy & C-AUC & C-GAUC & A-AUC & A-GAUC & O-AUC & O-GAUC \\
\midrule
\rowcolor{orange!5}
V1 & Causal Mask                                        & 0.839 & 0.836 & 0.876 & 0.830 & 0.911 & 0.838 \\
\rowcolor{orange!5}
V2 & Bi-Directional, CIS Inter-Invisible             & 0.860 & 0.859 & 0.903 & 0.848 & 0.920 & 0.847 \\
\rowcolor{orange!5}
V3 & Bi-Directional, CIS Inter-Visible               & 0.881 & 0.879 & 0.918 & 0.857 & 0.937 & 0.854 \\
\rowcolor{orange!15}
V4 & V3 + 1-Step Reasoning, Multi-Task on Last Step     & 0.890 & 0.889 & 0.931 & 0.871 & 0.946 & 0.867 \\
\rowcolor{orange!15}
V5 & V3 + 2-Step Reasoning, Multi-Task on Last Step     & 0.893 & 0.894 & 0.936 & 0.876 & 0.948 & 0.869 \\
\rowcolor{orange!15}
V6 & V3 + 3-Step Reasoning, Multi-Task on Last Step     & 0.906 & 0.902 & 0.946 & 0.889 & 0.957 & 0.881 \\
\rowcolor{orange!30}
V7 & V3 + 3-Step Reasoning, Progressive Multi-Task            & \textbf{0.911} & \textbf{0.909} & \textbf{0.952} & \textbf{0.897} & \textbf{0.963} & \textbf{0.886} \\
\bottomrule
\end{tabular}
}
\end{table}

We systematically examine the impact of different training strategies on both retrieval and ranking tasks to validate our progressive multi-task learning framework. The results are demonstrated in Table~\ref{tab:ablation_training_retrieval} and Table~\ref{tab:ablation_training_ranking}, respectively.

Starting from a causal attention baseline (V1), bidirectional attention (V2) effectively unlocks substantial gains across both tasks, improving retrieval R@100 from 0.464 to 0.470 and ranking C-AUC from 0.839 to 0.860. This validates our architecture design leveraging the bidirectional attention nature of recommendation tasks, where full context attention provides more comprehensive representation information. For ranking tasks, enabling candidate inter-visibility (V3) delivers another major boost, with C-AUC jumping from 0.860 to 0.881. This substantial improvement confirms that our candidate-aware framework enables the rich comparative reasoning essential for accurate ranking, allowing candidates to attend to each other rather than being processed in isolation.

The introduction of our block-wise reasoning mechanism demonstrates consistent and cumulative performance gains across both tasks. In retrieval, extending from direction decision-making to single-step reasoning with click prediction (V3) achieves notable improvements (R@100: 0.470 to 0.490), while multi-task learning on the final step (V4) provides additional gains (R@100: 0.495). Increasing reasoning depth yields expected progressive benefits: two-step reasoning (V5) improves retrieval R@100 to 0.510, while three-step reasoning in ranking (V6) achieves C-AUC of 0.906. These results establish that our reasoning framework enables increasingly sophisticated preference modeling, where each additional reasoning step meaningfully contributes to performance by capturing more nuanced user behavioral patterns.

Our progressive multi-task training consistently outperforms single-embedding multi-task learning at final step across both tasks. The key advantage lies in distributing different tasks across multiple reasoning steps rather than concentrating all supervision on a single final embedding. Progressive training surpasses the single-embedding approach with notable margins, improving R@100 from 0.510 (V5) to 0.517 (V6) in retrieval and C-AUC from 0.906 (V6) to 0.911 (V7) in ranking. This design prevents gradient conflicts that arise when multiple tasks compete for optimization at the one representation. Instead, each reasoning step serves as a specialized read-out token that helps the model extract task-specific information, effectively decoupling gradient flows across different objectives. The results also reveal task-specific optimization patterns: retrieval benefits from focused two-step reasoning aligned with the impression-click hierarchy, while ranking achieves optimal performance with three-step reasoning capturing the full conversion funnel, demonstrating how our progressive framework naturally adapts to different task complexities.

\subsection{Scaling Analysis}
In this section, we provide more in-depth analyses to better understand the effectiveness of OnePiece.

\subsubsection{Training Data Scaling}

\begin{figure}[t]
    \centering
    \includegraphics[width=\linewidth]{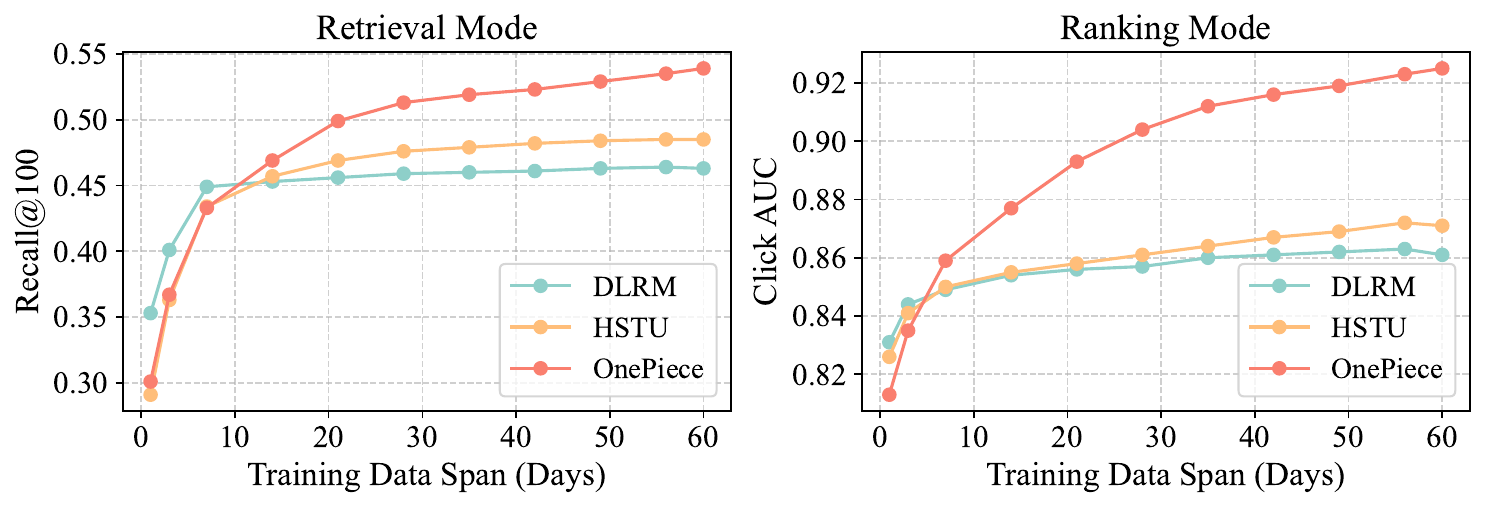}
    \caption{Training convergence curves of different models on retrieval and ranking tasks.}
    \label{fig:training_convergence}
\end{figure}

To analyze how performance scales with increasing training data, we report Recall@100 for retrieval and AUC for ranking over different training spans (\textit{i.e.}, the number of consecutive log days used for training), covering up to $60$ days of data. 
To ensure a fair comparison, we carefully match the model size (\textit{i.e.}, the total number of trainable parameters) across DLRM, HSTU, and our proposed OnePiece. 
As shown in Figure~\ref{fig:training_convergence}, OnePiece already surpasses both baselines after only $7$--$10$ days of training, demonstrating superior data efficiency attributable to its context-aware design and multi-step reasoning architecture. 
With longer training spans, DLRM and HSTU quickly converge to a plateau, whereas OnePiece continues to improve with a widening margin. 
By day 60, the performance gap between OnePiece and baseline methods (DLRM and HSTU) becomes pronounced, with the baseline models exhibiting convergence behavior while OnePiece continues to demonstrate improvement potential. This suggests that OnePiece has not yet reached full convergence and can achieve further performance gains with additional training data, indicating superior scaling capabilities.
These results highlight OnePiece’s stronger modeling capacity and its ability to effectively exploit richer behavioral supervision from extended time horizons. 
The trend is consistent across both retrieval and ranking tasks, and the training curves of OnePiece exhibit smooth and stable growth without noticeable fluctuations, indicating robust optimization under progressive multi-task supervision. 
Overall, OnePiece not only achieves higher sample efficiency but also scales more effectively as additional training data becomes available.

\subsubsection{Reasoning Scaling}
\begin{table}[t]
\centering
\caption{Effect of block size on ranking performance of OnePiece. The training data span is 60 days for this experiment.}
\label{tab:block_size_scaling}
\begin{tabular}{c cccccc}
\toprule
Block Size & C-AUC & C-GAUC & A-AUC & A-GAUC & O-AUC & O-GAUC \\
\midrule
$M=C=1$  & $0.885$ & $0.881$ & $0.923$ & $0.861$ & $0.947$ & $0.871$ \\
$M=C=4$  & $0.913$ & $0.911$ & $0.951$ & $0.896$ & $0.961$ & $0.885$ \\
$M=C=8$  & $0.920$ & $0.918$ & $0.956$ & $0.899$ & $0.964$ & $0.887$ \\
$M=C=12$ & $0.927$ & $0.923$ & $0.958$ & $0.903$ & $0.969$ & $0.893$ \\
\bottomrule
\end{tabular}
\end{table}

We investigate the impact of reasoning block size $M$ (\textit{i.e.}, the number of candidate items considered in ranking mode\footnote{Note that we only conduct block size scaling experiments in the ranking stage. 
In retrieval, the candidate items are not fed into the recommendation model, and each block contains only two global tokens (a user token capturing personalization and a query token capturing relevance). This design leaves little room for block-size ablation in retrieval mode.}) on the ranking performance of OnePiece. As shown in Table~\ref{tab:block_size_scaling}, increasing $M$ from $1$ to $12$ yields consistent improvements across all evaluated metrics. The largest performance gain appears when scaling from $M=1$ to $M=4$, since pointwise modeling ($M=1$) lacks cross-sample comparisons, while grouping candidates into blocks enables the reasoning mechanism to contrast preferences more effectively. As block size continues to increase, the improvements become smaller yet remain positive, indicating diminishing returns. This effect likely arises because overly large blocks overload the reasoning medium with redundant information, saturating its representational capacity. Overall, these findings reveal a trade-off between expanding reasoning bandwidth and avoiding information redundancy, underscoring the importance of selecting an appropriate block size to maximize the effectiveness of block-wise reasoning.

\section{Online A/B Testing}
In this section, we conduct large-scale online A/B testing on Shopee’s production search system to further evaluate the real-world effectiveness of OnePiece.

\subsection{Experimental Settings}
\subsubsection{Online Inference Details}

\textbf{Retrieval Stage.}
For retrieval, we conduct offline training to generate vector representations for the entire item pool, constructing an Approximate Nearest Neighbor (ANN)~\citep{indyk1998approximate} index to support efficient online retrieval. Specifically, we employ the Hierarchical Navigable Small World (HNSW)~\citep{malkov2018efficient} algorithm. 

\noindent \textbf{Ranking Stage.\footnote{Due to resource constraints in online deployment, the production version of OnePiece ranking model is a downgraded variant where the block size is reduced to $M{=}1$ and the number of item side-information features is significantly reduced.}}
For ranking, we adopt the following score fusion strategy to integrate outputs from different tasks:
\begin{equation}
\label{eq:score_fusion}
p_{\text{final}} = \alpha \cdot p_{\text{ctr}}^{a} \cdot p_{\text{ctcvr}}^{b} + \beta \cdot p_{\text{ctr}}^{a} \cdot p_{\text{ctcvr}}^{b} \cdot \text{price} + \gamma \cdot p_{\text{ctr}} \cdot \text{ecpm},
\end{equation}
where $\alpha$, $\beta$, and $\gamma$ are hyperparameters controlling the importance weights of respective components, enabling the balance between user experience and business revenue. $p_{\text{ctr}}$ and $p_{\text{ctcvr}}$ represent the click-through rate and click-to-conversion rate predicted by OnePiece's final reasoning step, corresponding to the logits of the click and order tasks respectively. Parameters $a$ and $b$ modulate the influence of each task in the final ranking, while $\text{price}$ denotes the item's price information and $\text{ecpm}$ represents the item's advertising value component.

\subsubsection{Evaluation Protocol}

In industrial personalized ranking systems, candidate generation typically follows a cascade paradigm consisting of multiple stages (\textit{e.g.}, retrieval, pre-ranking, ranking). The retrieval stage usually generates a large set of candidate items through multiple parallel recall strategies, each capturing different facets of user preference to ensure diversified coverage, while the subsequent pre-ranking and ranking stages aim to produce the final ranked list with fine-grained preference estimation.  As an initial online exploration, we allocate $10\%$ of traffic for all A/B testing experiments under the following settings:
\begin{itemize}
    \item For retrieval evaluation, we replace one of the parallel recall strategies with our proposed OnePiece retrieval model, specifically replacing the User-to-Item (U2I) recall route originally implemented in DLRM.
   \item For ranking evaluation, we substitute the DLRM model in the pre-ranking stage with our proposed OnePiece ranking model. 
\end{itemize}
These designs allow us to isolate the contribution of OnePiece at different modes (\textit{i.e.}, retrieval and ranking) within the industrial cascade ranking systems.

\subsubsection{Evaluation Metrics}
We report the following business and user engagement indicators tracked in our online experiment: 
\vspace{-0.3cm}
\begin{itemize}
    \item \textbf{GMV/UU} (Gross Merchandise Volume per Unique User): Average transaction value contributed per user.  

    \item \textbf{GMV(99.5\%)/UU}: GMV per user excluding the top $0.5\%$ high-value orders, reflecting stable contributions from regular transactions.  

    \item  \textbf{AR/UU} (Advertising Revenue per Unique User): Average advertising revenue generated per unique user, reflecting conversion efficiency from ads exposure.

    \item \textbf{Order/UU}: Average number of orders per user, capturing transaction frequency.  

    \item \textbf{Paid Order/UU}: Average number of successfully paid orders per user, counting only completed purchases without refund.   
    \item \textbf{CTR} (Click-Through-Rate): The ratio of clicked impressions to total impressions, reflecting the effectiveness of ranked results in attracting user engagement.
    
    \item \textbf{CTCVR} (Click-to-Conversion Rate): The ratio of successful conversions to total clicks, measuring the effectiveness of transforming user engagement into completed transactions.

    \item \textbf{Buyer}: Proportion of unique users who placed at least one order.  

    \item \textbf{Bad Query Rate}: The percentage of queries for which human evaluators judge the recommended content as irrelevant to the query, serving as an inverse measure of recommendation accuracy and user satisfaction.
\end{itemize}

\begin{table}[t]
\centering
\caption{Results of online A/B testing for the retrieval mode of OnePiece. Improvements are reported in relative percentage changes over the DLRM baseline.}
\label{tab:online_ab_recall}
\begin{tabular}{ccccccc}
\toprule
 GMV/UU & GMV(99.5\%)/UU & Order/UU & Paid Order/UU & CTCVR & Buyer & Bad Query Rate  \\
\midrule
$+1.08\%$ & $+0.91\%$ & $+0.71\%$ & $+0.98\%$ & $+0.66\%$ & $+0.41\%$ & $-0.17\%$\\
\bottomrule
\end{tabular}
\end{table}

\begin{table}[t]
\centering
\caption{Results of online A/B testing for the ranking mode of OnePiece. Improvements are reported in relative percentage changes over the DLRM baseline.}
\label{tab:online_ab_rank}
\begin{tabular}{ccccccc}
\toprule
 GMV/UU & GMV(99.5\%)/UU & AR/UU & Order/UU & Buyer & CTR & Bad Query Rate \\
\midrule
$+1.12\%$ & $+0.65\%$ & $+2.90\%$ & $+0.08\%$ & $+0.08\%$ & $+0.29\%$ & $+0.21\%$ \\
\bottomrule
\end{tabular}
\end{table}

\subsection{Overall Performance}

\textbf{Retrieval Mode.} Table~\ref{tab:online_ab_recall} reports the detailed results of our online A/B testing for the retrieval mode of OnePiece. We observe consistent improvements across all key business metrics. GMV/UU increases by $+1.08\%$, and even after removing the top $0.5\%$ of high-value orders, GMV(99.5\%)/UU still improves by $+0.91\%$, suggesting that the gain is not driven solely by occasional big-ticket purchases but reflects stable contributions from regular transactions. Order/UU rises by $+0.71\%$, while Paid Order/UU grows even faster ($+0.98\%$) than Order/UU, indicating both higher conversion and a reduced refund rate. Buyer expands by $+0.41\%$, meaning that more unique users were successfully converted. At the same time, CTCVR improves by $+0.66\%$, reflecting a higher end-to-end conversion rate from exposure to conversion. Importantly, the Bad Query Rate decreases by $0.17\%$, which indicates better query relevance and improved user experience. Unlike prior personalized recall strategies that often boosted GMV at the expense of relevance, OnePiece achieves balanced improvements, simultaneously enhancing personalization, relevance, and overall transaction stability.

\textbf{Ranking Mode.} Table~\ref{tab:online_ab_rank} summarizes the online A/B testing results of deploying OnePiece in the ranking stage. We observe consistent lifts across major business metrics: GMV/UU improves by $+1.12\%$, and advertising revenue (AR/UU) shows a substantial boost of $+2.9\%$. In contrast, Order/UU increases only marginally ($+0.08\%$), which aligns with our design that translates order-related utility into GMV and advertising gains via the score fusion function (Eq.~\ref{eq:score_fusion}). Buyer also rises slightly by $+0.08\%$, indicating broader coverage of converted users. On engagement metrics, CTR improves by $+0.29\%$, suggesting stronger attractiveness of ranked results. Meanwhile, the Bad Query Rate increases by $+0.21\%$, likely due to more advertising slots introducing items less relevant to user interests. However, this minor increase is outweighed by the substantial revenue gains. Overall, these results demonstrate that OnePiece ranking not only strengthens core business metrics but also achieves a practical trade-off between user experience and business objectives in large-scale industrial ranking systems.

\subsection{Recall Coverage and Exclusive Contribution}\label{sec:recall_coverage}

\begin{table}[t]
\centering
\caption{Comparison of overlap coverage between DLRM and OnePiece with respect to other recall strategies.
For each recall route $R$, the recall coverage is computed as 
$\text{Coverage}(R) = \frac{|\,\text{Exposure}_{\text{U2I}} \cap \text{Exposure}_{R}\,|}{|\,\text{Exposure}_{R}\,|}$. 
Here, $\text{Exposure}_{\text{U2I}}$ denotes the exposure set attributed to the U2I model (DLRM or OnePiece), and $\text{Exposure}_{R}$ denotes the exposure set of another recall route. 
\textbf{STR1}: sparse text recall with user-input keywords; 
\textbf{STR2}: sparse text recall with rewritten keywords; 
\textbf{Swing I2I}: graph-based item-to-item personalized recall; 
\textbf{KPop}: popularity-based recall under keywords; 
\textbf{S2I}: semantic vector-to-item recall. 
Relative improvements of OnePiece over DLRM are shown in red.}
\label{tab:exp_recall_coverage}
\resizebox{1\linewidth}{!}{
\begin{tabular}{lccccc}
\toprule
Recall Route & STR1 & STR2 & Swing I2I & KPop & S2I \\
\midrule
DLRM & 37.3\% & 31.3\% & 57.9\% & 62.5\% & 47.6\% \\
OnePiece & 66.2\% (\textcolor{red}{+77.6\%}) 
          & 64.4\% (\textcolor{red}{+105.8\%}) 
          & 76.8\% (\textcolor{red}{+32.6\%}) 
          & 77.2\% (\textcolor{red}{+23.5\%}) 
          & 67.8\% (\textcolor{red}{+42.4\%}) \\
\bottomrule
\end{tabular}
}
\end{table}

To further evaluate the effectiveness of OnePiece in the retrieval stage, we analyze the overlap between the exposure items contributed by OnePiece and those from other recall routes (\textit{i.e.}, recall coverage), in comparison with the traditional DLRM baseline. As shown in Table~\ref{tab:exp_recall_coverage}, OnePiece consistently achieves higher recall coverage across all routes. In sparse text recall, it improves STR1 coverage from $37.3\%$ to $66.2\%$ ($+77.6\%$) and STR2 coverage from $31.3\%$ to $64.4\%$ ($+105.8\%$). Similar gains are observed in graph-based Swing I2I ($+32.6\%$) and keyword-popularity recall KPop ($+23.5\%$), while semantic recall S2I also increases significantly from $47.6\%$ to $67.8\%$ ($+42.4\%$). These consistent improvements demonstrate that, compared to DLRM, OnePiece achieves substantial coverage gains over all other recall routes, revealing its strong potential to replace multiple specialized recall strategies with a single unified model. We believe that, with carefully designed context engineering, OnePiece can effectively balance personalization (Swing I2I), popularity (KPop), and relevance (STR1 and STR2), thereby moving closer to a ``one model serves all purposes'' paradigm for industrial-scale recall.

\begin{wrapfigure}{r}{0.4\linewidth}
    \centering
    \vspace{-0.5cm}
    \includegraphics[width=\linewidth]{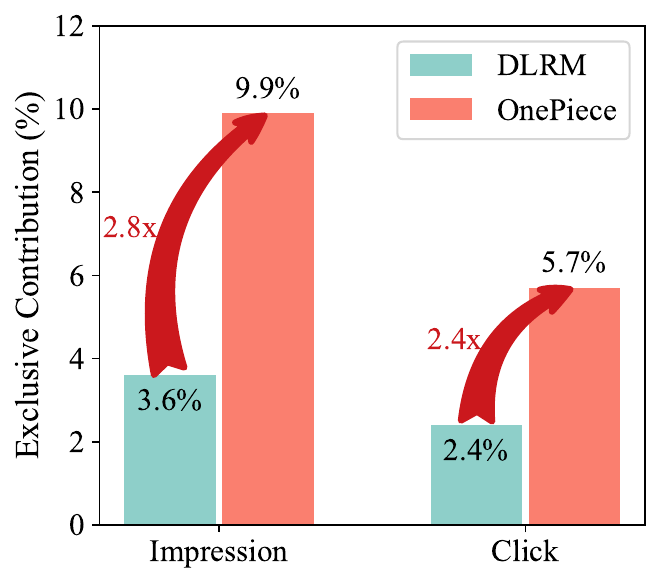}
    \caption{Exclusive contribution of OnePiece in the retrieval stage.}
    \vspace{-1em}
    \label{fig:exclusive_recall_contribution}
\end{wrapfigure}
To further examine the independent value of OnePiece, we compare the exclusive contribution of OnePiece and DLRM in terms of impressions and clicks, as shown in Figure~\ref{fig:exclusive_recall_contribution}. OnePiece demonstrates a substantial increase in unique contribution: exclusive impression share rises from $3.6\%$ to $9.9\%$ (2.8x), while exclusive click share grows from $2.4\%$ to $5.7\%$ (2.4x). These results indicate that OnePiece is not only capable of covering the exposure of other recall routes but also contributes significantly more novel impressions and clicks that are not captured by traditional DLRM-based recall. In other words, the new recall route of OnePiece nearly doubles the independent value over traditional DLRM, showcasing its effectiveness in enhancing overall recall performance.

\subsection{Efficiency Analysis}

\begin{table}[h]
\centering
\caption{Computational efficiency comparison between DLRM and OnePiece under both retrieval and ranking modes, evaluated on a single NVIDIA A30 GPU. Here, \textbf{MFU} denotes \textit{Model FLOPs Utilization}, reflecting the ratio of achieved FLOPs to the theoretical peak, while \textbf{MU} denotes \textit{Memory Utilization}, measured as the percentage of GPU memory capacity occupied during inference. The arrows indicate the preferred direction for each metric: ``$\uparrow$'' for higher is better while ``$\downarrow$'' for lower is better.}
\label{tab:efficiency}
\begin{tabular}{lccc}
\toprule
\multicolumn{4}{c}{\textbf{Retrieval Mode}} \\
\textbf{Method} & \textbf{Infer. Time$\downarrow$} & \textbf{MFU$\uparrow$} & \textbf{MU$\uparrow$} \\
\midrule
DLRM & 40ms/request & 35\% & 30\% \\
OnePiece & 30ms/request (\textcolor{red}{-25\%}) & 80\% (\textcolor{red}{+129\%}) & 50\% (\textcolor{red}{+67\%}) \\
\midrule
\multicolumn{4}{c}{\textbf{Ranking Mode (batch size=128, KV-Cache enabled)}} \\
\textbf{Method} & \textbf{Infer. Time$\downarrow$} & \textbf{MFU$\uparrow$} & \textbf{MU$\uparrow$} \\
\midrule
DLRM & 109ms/batch & 23\% & 29\% \\
OnePiece (M=1) & 110ms/batch (+0.9\%) & 67\% (\textcolor{red}{+191\%}) & 38\% (\textcolor{red}{+31\%}) \\
OnePiece (M=4) & 112ms/batch (+2.8\%) & - & - \\
OnePiece (M=8) & 115ms/batch (+5.5\%) & - & - \\
OnePiece (M=12) & 120ms/batch (+10.1\%) & - & - \\
\bottomrule
\end{tabular}
\end{table}

Based on the computational efficiency analysis presented in Table~\ref{tab:efficiency}, OnePiece exhibits superior hardware utilization and execution performance characteristics that support its viability for large-scale industrial deployment across both retrieval and ranking modes.

\textbf{Enhanced Hardware Utilization in Retrieval Mode.}
OnePiece delivers significant efficiency gains over DLRM in retrieval tasks: a \textbf{25\% reduction in inference time} (30ms vs.\ 40ms per request) while simultaneously achieving dramatic improvements in resource utilization. The \textbf{129\% increase in Model FLOPs Utilization} (from 35\% to 80\%) indicates that OnePiece's unified Transformer architecture exhibits superior compatibility with modern GPU parallelization paradigms, effectively leveraging tensor computation units that remain underutilized in traditional embedding-heavy architectures like DLRM. The concurrent 67\% increase in memory utilization (from 30\% to 50\%) represents better resource exploitation, demonstrating more effective utilization of available hardware capacity while maintaining faster inference speeds. This efficiency improvement suggests that OnePiece's architectural unification eliminates the computational bottlenecks inherent in DLRM's heterogeneous component interactions, enabling more streamlined data flow and reduced memory transfer overhead. The superior hardware utilization translates directly to  reduced operational costs, critical factors for large-scale industrial deployment where processing millions of requests daily.

\textbf{Controlled Computational Scaling in Ranking Mode.}
The ranking mode evaluation reveals OnePiece's capacity for efficient scaling with increased reasoning complexity. While OnePiece incurs modest overhead relative to DLRM (109ms baseline), the scaling behavior with block size exhibits desirable sub-linear characteristics: inference time increases from 110ms at $M{=}1$ to 120ms at $M{=}12$, representing only a \textbf{10.1\% overhead for $12\times$ reasoning capacity expansion}. The progressive overhead pattern (0.9\% $\rightarrow$ 2.8\% $\rightarrow$ 5.5\% $\rightarrow$ 10.1\%) demonstrates that the block-wise reasoning mechanism achieves efficient computational amortization, where each additional reasoning block incurs diminishing marginal cost.
This controlled scaling translates into a highly favorable efficiency-performance trade-off: as demonstrated in Table~\ref{tab:block_size_scaling}, C-AUC improves from 0.885 at $M=1$ to 0.927 at $M=12$, representing a significant \textbf{4.7\%} relative improvement, uncovering the potential for reasoning scaling in our unified framework.
Particularly noteworthy is the dramatic \textbf{191\% improvement in MFU} (from 23\% to 67\%) achieved even at $M{=}1$, indicating that OnePiece's unified architecture aligns with GPU computational paradigms regardless of reasoning complexity. 
This efficiency gain stems from the infrastructure trick: the \textbf{KV-Caching} mechanism enables efficient batch processing while maintaining this superior hardware utilization, demonstrating the effectiveness in handling high-throughput production workloads.
These findings establish OnePiece as a practical solution for production systems, offering configurable reasoning depth that enables flexible trade-offs between computational efficiency and model performance.

%% file: section/related_work.tex
\section{Related Work}

\textbf{Context Engineering and Reasoning Enhancement in LLMs.}
The Transformer architecture~\citep{vaswani2017attention} has fundamentally transformed artificial intelligence across multiple domains~\citep{brown2020language,dosovitskiy2020image,wang2017tacotron} and culminating in ChatGPT's demonstration of emergent general-purpose capabilities in large language models (LLMs). Beyond established scaling laws~\citep{zhao2023survey,isik2024scaling,chung2024scaling}, current research has identified two orthogonal optimization pathways for maximizing LLM utility: \textbf{Context Engineering} and \textbf{Reasoning Enhancement}. In specific, context engineering encompasses the evolution from basic prompt engineering~\citep{marvin2023prompt,chen2023unleashing,cain2024prompting} toward comprehensive context architecture~\citep{mei2025survey,amatriain2024prompt,velasquez2023prompt} that orchestrate prompts, external knowledge bases, persistent memory, and interaction histories into input context to effectively steer reasoning and generation. The reasoning enhancement pathway advances computational reasoning through sophisticated methodologies including chain-of-thought prompting~\citep{wei2022chain,zhang2022automatic,diao2023active}, structured tree-based~\citep{yao2023tree,long2023large,cao2023probabilistic} and graph-based reasoning~\citep{jin2024graph,yao2023beyond,besta2024graph}, and iterative self-refinement mechanisms~\citep{yan2024mirror,madaan2023self,ranaldi2024self} that augment latent multi-step reasoning processes~\citep{hao2024training,zhu2025survey,su2025token}, enabling logical decomposition and strategic planning that transcends statistical pattern matching. The convergence of these complementary research directions transforms LLMs from probabilistic sequence generators into adaptive reasoning architectures capable of complex problem-solving and knowledge integration.

\textbf{Context Engineering and Reasoning Enhancement in Ranking Systems.}
The evolution of ranking systems has progressively integrated advances in language modeling, with early explorations mainly focusing on architectural adaptations by directly applying Transformer-based models like SASRec~\citep{kang2018self} and BERT4Rec~\citep{sun2019bert4rec}.
Inspired by advances in large language models (LLMs), recent research has crystallized into two promising research directions that mirror established LLM optimization pathways.
First, context engineering-based approaches focus on developing sophisticated user sequence representations through dynamic behavior modeling and retrieval-augmented methods, with representative works~\citep{lin2024rella,li2024rat} transitioning from conventional recency-based sequence modeling toward semantic behavior retrieval algorithms that extract contextually relevant historical interactions, demonstrating enhanced robustness and improved cold-start prediction accuracy. 
Second, reasoning enhancement-based methods have bifurcated into explicit and implicit paradigms~\citep{yu2025thinkrec,zhang2025slow,wang2024can}, where explicit reasoning approaches~\citep{fang2025reason4rec,gu2025r4ec} employ multi-expert pipelines and reflection-refinement mechanisms to generate explainable intermediate reasoning at the cost of computational overhead, while implicit reasoning methods~\citep{tang2025think,liu2025lares, zhang2025reinforced} leverage continuous hidden state autoregression and depth-recurrent architectures to achieve enhanced reasoning capabilities without explicit text generation requirements, offering computational efficiency while maintaining reasoning depth through specialized alignment objectives.

While existing efforts have explored various LLM-inspired techniques for ranking systems, they lack a principled approach that captures the fundamental design principles driving LLM success. To this end, we propose OnePiece, which addresses this gap by establishing a unified architectural paradigm that comprehensively adapts the core design philosophy of LLMs to ranking systems.

%% file: section/appendix.tex
\section{Offline Dataset Construction}\label{app:offline_dataset_cons}

\textbf{Retrieval Stage.}
The objective of the retrieval task is to retrieve items that users may potentially interact with in the future. Therefore, we focus on learning impression and click objectives. For each sample, we first filter out session request data where users did not exhibit click behaviors from the online data. Then, we construct the training samples as follows: (1) the $m$ clicked items serve as positive samples for both impression and click tasks; (2) the $n$ exposed but unclicked items serve as positive samples for the impression task while simultaneously acting as negative samples for the click task; (3) we further sample $k$ items from unexposed items within the top-500 results from the ranking stage as additional negative samples; (4) we sample $l$ items from the same category as the clicked items, which serve as hard negative samples to enhance model convergence and mitigate homogeneous recommendation risks. The specific values of $m$, $n$, $k$, and $l$ are determined based on domain expert experience and empirical validation.

\noindent \textbf{Ranking Stage.}
As the downstream stage following retrieval, the ranking task requires further refinement of user interests and preferences to more accurately compute relevance scores for candidate items. We filter session request data to retain only those with click behaviors. For training sample construction, we use task-specific interaction types (\textit{i.e.}, impression, click, add-to-cart, order) as positive samples, while interactions from preceding tasks in the conversion funnel serve as negative samples for each respective task (\textit{e.g.}, for the order prediction task, exposed, clicked, and add-to-cart items that were not purchased serve as negative samples). Additionally, analogous to the retrieval stage, we enhance the negative sample pool by randomly sampling items from the top-500 ranking results that were not exposed to users, which serve as augmented hard negative samples to improve model performance.

\section{Context Engineering Details}

For the input components, namely Interaction History (IH), Preference Anchor (PA), Situational Descriptors (SD), and Candidate Item Set (CIS), we detail the construction as follows. As CIS involves straightforward item features, here, we mainly focus on the other three components.

\textbf{Interaction History (IH):} We utilize users' click sequences, shopping cart sequences, and purchase sequences from the most recent month, chronologically sorted and merged. Each item in the sequence encompasses features including item ID, category, shop, and other relevant attributes.

\textbf{Preference Anchor (PA):} We aggregate users' query-associated top-k exposures, top-k clicks, and top-k purchases, utilizing [BOS] and [EOS] tokens to separate different sequence types. Unlike the IH component, we employ sequence concatenation rather than mixing approaches.

\textbf{Situational Descriptors (SD):} This component covers user profiles and query-related contextual information. We introduce two specialized tokens: User Token (integrating user ID, age, location, and other demographic features) and Query Token (incorporating query ID, textual content, query popularity, and related information).

\section{Attention Visualization Analysis}

\begin{figure}[H]
    \centering
    {
        \captionsetup{justification=centering}
        \begin{subfigure}{\linewidth}
            \centering
            \includegraphics[width=\linewidth]{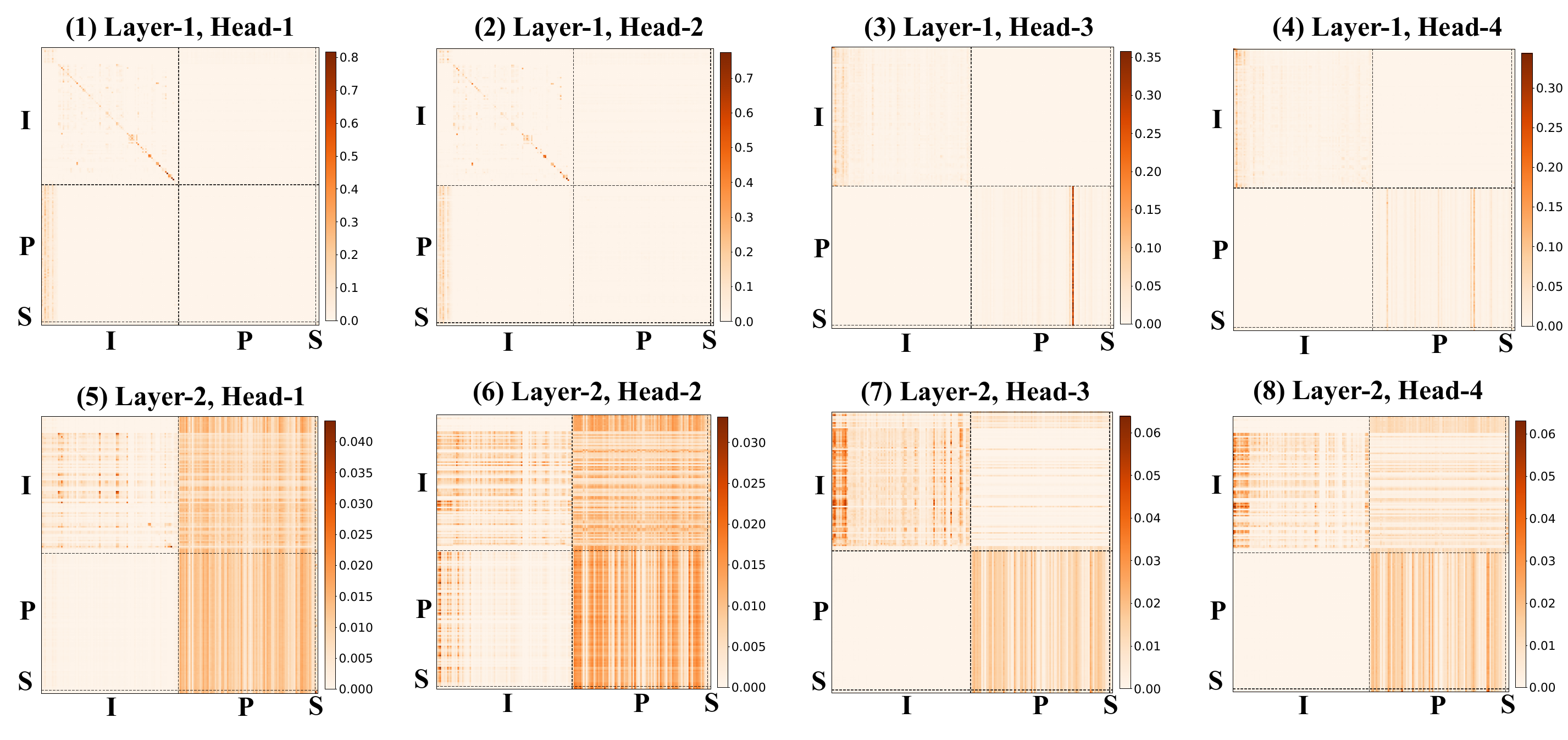}
            \vspace{-1em}
            \caption{Case Studies on OnePiece Attention Analysis (Retrieval Mode).}
            \label{fig:retrieval_attention_map}
        \end{subfigure}
    }
    \vspace{2em}
    {
        \captionsetup{justification=centering}
        \begin{subfigure}{\linewidth}
            \centering
            \includegraphics[width=\linewidth]{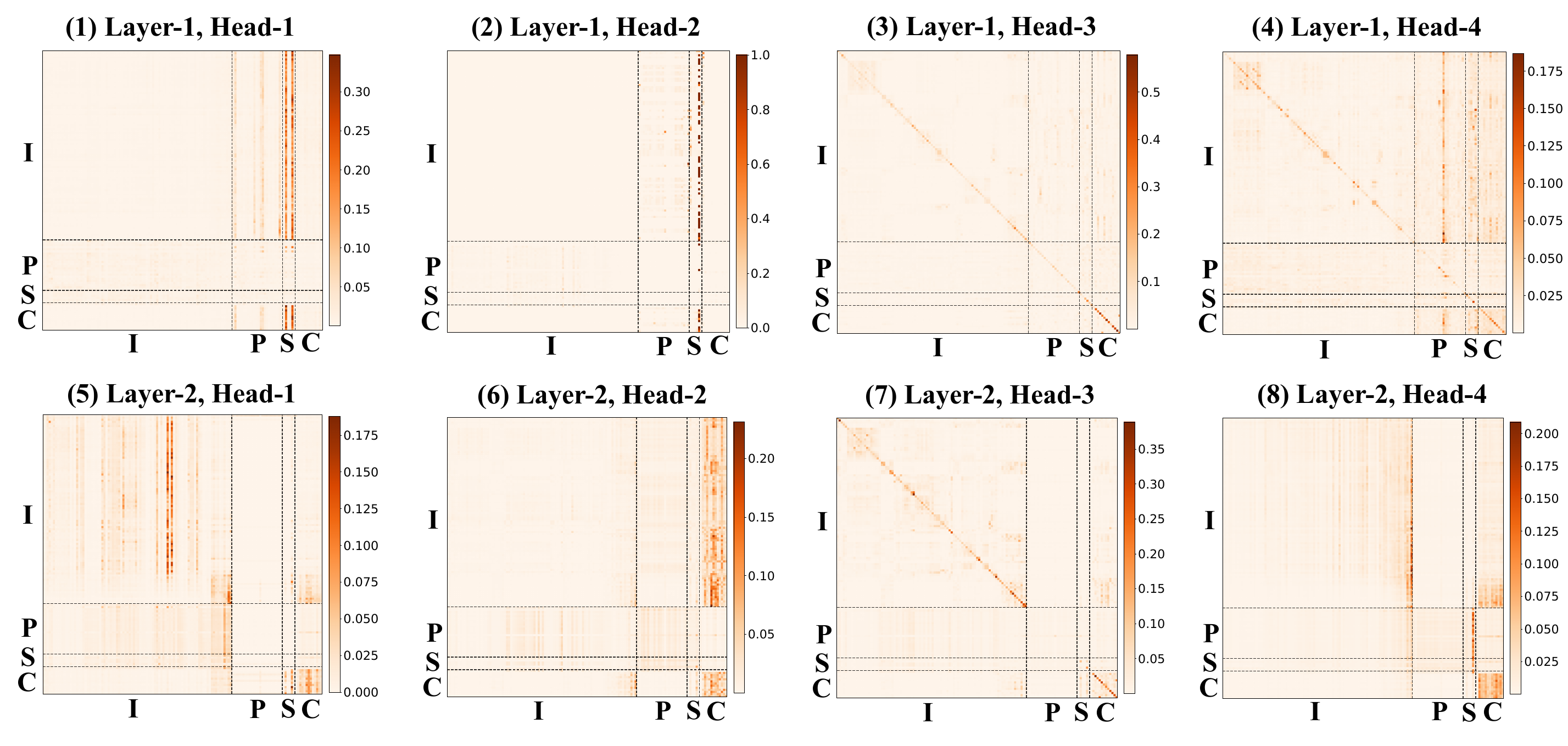}
            \vspace{-1em}
            \caption{Case Studies on OnePiece Attention Analysis (Ranking Mode).}
            \label{fig:ranking_attention_map}
        \end{subfigure}
    }
    \vspace{-2em}
    \caption{OnePiece Attention Analysis in Different Modes. The attention maps visualize the attention weights between different input components: Interaction History (I), Preference Anchor (P), Situational Descriptor (S), and Candidate Item Set (C). In these visualizations, the \textbf{y-axis} represents the Query, while the \textbf{x-axis} represents the Key-Value, corresponding to the attention weight matrix commonly used in Transformer-like architectures.}
    \label{fig:attention_analysis_combined}
\end{figure}

\paragraph{Attention Analysis of Context Input.}
Figure~\ref{fig:attention_analysis_combined} illustrates how our unified token sequence (IH, PA, SD, CIS) facilitates structured preference reasoning across both retrieval and ranking stages. Across modes, we observe two consistent trends. First, \textit{layer-wise evolution} is evident: Layer-1 heads (\textit{e.g.}, Figure~\ref{fig:attention_analysis_combined}(a)-1--4 and Figure~\ref{fig:attention_analysis_combined}(b)-1--3) primarily form concentrated or diagonal attention bands, highlighting localized sequential processing within IH tokens or short-span links between SD and CIS. In contrast, Layer-2 heads (Figure~\ref{fig:attention_analysis_combined}(a)-5--8; Figure~\ref{fig:attention_analysis_combined}(b)-5--8) develop multi-region attention patterns that simultaneously connect multiple token groups, suggesting a transition from localized to global integration. Second, \textit{head-level specialization} emerges clearly. Within the same layer, different heads adopt complementary strategies: some emphasize intra-component coherence (\textit{e.g.}, IH--IH diagonals in Figure~\ref{fig:attention_analysis_combined}(b)-7), while others allocate strong weights to cross-component flows (\textit{e.g.}, SD$\to$IH in Figure~\ref{fig:attention_analysis_combined}(a)-5,6; CIS$\to$IH in Figure~\ref{fig:attention_analysis_combined}(b)-6). Together, these patterns validate that the model learns hierarchical and diversified reasoning strategies rather than redundantly replicating attention across different heads of OnePiece.

Beyond these shared behaviors, each mode exhibits distinctive characteristics aligned with its task. In retrieval mode (Figure~\ref{fig:attention_analysis_combined}(a)), the three-token design (IH, PA, SD) fosters \textit{structured and compact cross-component attention}. For example, Head-1 and Head-2 at Layer-2 (Figure~\ref{fig:attention_analysis_combined}(a)-5,6) concentrate on linking IH with PA, highlighting how anchors guide long-term preference recall, while Head-4 (Figure~\ref{fig:attention_analysis_combined}(a)-8) reinforces SD$\to$IH connections, grounding retrieval in situational context. These interactions remain relatively localized, reflecting retrieval's objective of coarse-grained candidate filtering. By contrast, ranking mode (Figure~\ref{fig:attention_analysis_combined}(b)) introduces CIS tokens, fundamentally expanding the attention space to \textit{four-way interactions}. This is most pronounced in Head-2 and Head-4 at
Layer-2 (Figure~\ref{fig:attention_analysis_combined}(b)-6,8), where attention flows span IH, PA, SD, and CIS simultaneously, enabling joint evaluation of user preference signals against explicit candidate items. Importantly, we see \textit{functional differentiation}: IH tokens preserve temporal sequentiality (Figure~\ref{fig:attention_analysis_combined}(b)-7), while CIS tokens actively integrate with PA and SD to enable fine-grained candidate comparison (Figure~\ref{fig:attention_analysis_combined}(b)-4). These entangled attention maps underscore the ranking stage’s role in nuanced discrimination among candidates, complementing the retrieval stage’s more selective filtering focus.

\begin{figure}[t]
    \centering
    \includegraphics[width=\linewidth]{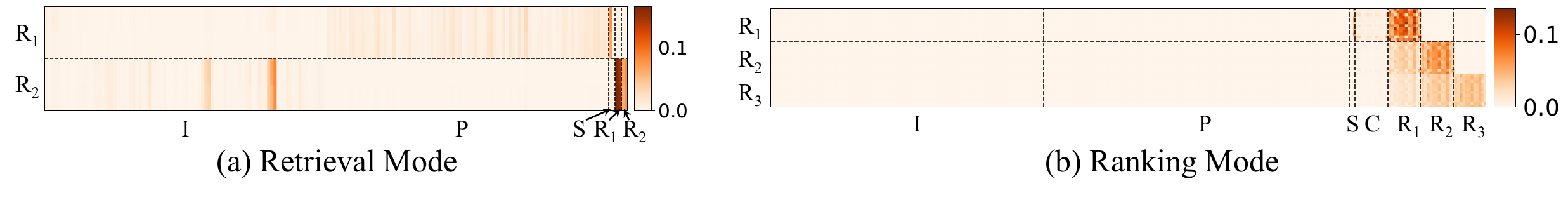}
    \caption{Attention visualization of multi-step block-wise reasoning in OnePiece. The heatmaps show attention weights between reasoning blocks (y-axis, as queries) and input components with previous reasoning outputs (x-axis, as keys and values) for \textbf{(a)} retrieval mode with two reasoning steps ($R_1$, $R_2$) and \textbf{(b)} ranking mode with three reasoning steps ($R_1$, $R_2$, $R_3$). Context Tokens include Interaction History (I), Preference Anchors (P), Situational Descriptors (S), and Candidate Items (C, ranking only).}
    \label{fig:reasoning_attention}
\end{figure}

\paragraph{Attention Analysis of Multi-Step Block-wise Reasoning.}
The attention patterns in Figure~\ref{fig:reasoning_attention} demonstrate how reasoning blocks progressively query different information sources to refine user representations. In \textbf{retrieval mode}, $R_1$ exhibits strong concentrated attention on situational descriptors (S) and moderate attention on preference anchors (P), with minimal attention to interaction history (I). This pattern indicates that initial reasoning prioritizes contextual and query-specific signals for user intent understanding. $R_2$ shows a pivotal shift, developing concentrated attention on specific regions within interaction history (I) while also incorporating information from the previous reasoning block $R_1$. This evolution from situational-preference focus to selective behavioral pattern recognition demonstrates how progressive reasoning enables the model to transition from broad contextual understanding to targeted sequential preference extraction.

In \textbf{ranking mode}, the three-step reasoning process ($R_1 \rightarrow R_2\rightarrow R_3$) reveals increasingly sophisticated attention integration with a notable hierarchical information enhancement pattern. As reasoning progresses, later blocks demonstrate stronger attention to more recent reasoning outputs, with $R_3$ showing pronounced attention to $R_2$ while exhibiting relatively weaker attention to $R_1$. This attention pattern suggests that each reasoning step progressively consolidates and refines information from previous steps, with more recent reasoning blocks containing higher-level abstractions that subsume earlier insights. The model effectively learns to prioritize the most refined representations rather than repeatedly accessing raw earlier outputs, indicating an efficient information compression mechanism where each reasoning step builds upon increasingly compressed preference understanding to achieve discriminative candidate evaluation.

%% file: main.bbl
\begin{thebibliography}{60}
\providecommand{\natexlab}[1]{#1}
\providecommand{\url}[1]{\texttt{#1}}
\expandafter\ifx\csname urlstyle\endcsname\relax
  \providecommand{\doi}[1]{doi: #1}\else
  \providecommand{\doi}{doi: \begingroup \urlstyle{rm}\Url}\fi

\bibitem[Ai et~al.(2019)Ai, Hill, Vishwanathan, and Croft]{ai2019zero}
Q.~Ai, D.~N. Hill, S.~Vishwanathan, and W.~B. Croft.
\newblock A zero attention model for personalized product search.
\newblock In \emph{Proceedings of the 28th ACM International Conference on Information and Knowledge Management}, pages 379--388, 2019.

\bibitem[Amatriain(2024)]{amatriain2024prompt}
X.~Amatriain.
\newblock Prompt design and engineering: Introduction and advanced methods.
\newblock \emph{arXiv preprint arXiv:2401.14423}, 2024.

\bibitem[Bengio et~al.(2009)Bengio, Louradour, Collobert, and Weston]{bengio2009curriculum}
Y.~Bengio, J.~Louradour, R.~Collobert, and J.~Weston.
\newblock Curriculum learning.
\newblock In \emph{Proceedings of the 26th annual international conference on machine learning}, pages 41--48, 2009.

\bibitem[Besta et~al.(2024)Besta, Blach, Kubicek, Gerstenberger, Podstawski, Gianinazzi, Gajda, Lehmann, Niewiadomski, Nyczyk, et~al.]{besta2024graph}
M.~Besta, N.~Blach, A.~Kubicek, R.~Gerstenberger, M.~Podstawski, L.~Gianinazzi, J.~Gajda, T.~Lehmann, H.~Niewiadomski, P.~Nyczyk, et~al.
\newblock Graph of thoughts: Solving elaborate problems with large language models.
\newblock In \emph{Proceedings of the AAAI conference on artificial intelligence}, volume~38, pages 17682--17690, 2024.

\bibitem[Brown et~al.(2020)Brown, Mann, Ryder, Subbiah, Kaplan, Dhariwal, Neelakantan, Shyam, Sastry, Askell, et~al.]{brown2020language}
T.~Brown, B.~Mann, N.~Ryder, M.~Subbiah, J.~D. Kaplan, P.~Dhariwal, A.~Neelakantan, P.~Shyam, G.~Sastry, A.~Askell, et~al.
\newblock Language models are few-shot learners.
\newblock \emph{Advances in neural information processing systems}, 33:\penalty0 1877--1901, 2020.

\bibitem[Cain(2024)]{cain2024prompting}
W.~Cain.
\newblock Prompting change: Exploring prompt engineering in large language model ai and its potential to transform education.
\newblock \emph{TechTrends}, 68\penalty0 (1):\penalty0 47--57, 2024.

\bibitem[Cao et~al.(2023)Cao, Zhang, Shi, Lv, Yao, Tian, Li, and Hou]{cao2023probabilistic}
S.~Cao, J.~Zhang, J.~Shi, X.~Lv, Z.~Yao, Q.~Tian, J.~Li, and L.~Hou.
\newblock Probabilistic tree-of-thought reasoning for answering knowledge-intensive complex questions.
\newblock \emph{arXiv preprint arXiv:2311.13982}, 2023.

\bibitem[Chen et~al.(2023)Chen, Zhang, Langren{\'e}, and Zhu]{chen2023unleashing}
B.~Chen, Z.~Zhang, N.~Langren{\'e}, and S.~Zhu.
\newblock Unleashing the potential of prompt engineering in large language models: a comprehensive review.
\newblock \emph{arXiv preprint arXiv:2310.14735}, 2023.

\bibitem[Chen et~al.(2025)Chen, Guo, Wang, Liang, Lv, Ma, Xiao, Xue, Zhang, Yang, et~al.]{chen2025onesearch}
B.~Chen, X.~Guo, S.~Wang, Z.~Liang, Y.~Lv, Y.~Ma, X.~Xiao, B.~Xue, X.~Zhang, Y.~Yang, et~al.
\newblock Onesearch: A preliminary exploration of the unified end-to-end generative framework for e-commerce search.
\newblock \emph{arXiv preprint arXiv:2509.03236}, 2025.

\bibitem[Chung et~al.(2024)Chung, Hou, Longpre, Zoph, Tay, Fedus, Li, Wang, Dehghani, Brahma, et~al.]{chung2024scaling}
H.~W. Chung, L.~Hou, S.~Longpre, B.~Zoph, Y.~Tay, W.~Fedus, Y.~Li, X.~Wang, M.~Dehghani, S.~Brahma, et~al.
\newblock Scaling instruction-finetuned language models.
\newblock \emph{Journal of Machine Learning Research}, 25\penalty0 (70):\penalty0 1--53, 2024.

\bibitem[Deng et~al.(2025)Deng, Wang, Cai, Ren, Hu, Ding, Luo, and Zhou]{deng2025onerec}
J.~Deng, S.~Wang, K.~Cai, L.~Ren, Q.~Hu, W.~Ding, Q.~Luo, and G.~Zhou.
\newblock Onerec: Unifying retrieve and rank with generative recommender and iterative preference alignment.
\newblock \emph{arXiv preprint arXiv:2502.18965}, 2025.

\bibitem[Diao et~al.(2023)Diao, Wang, Lin, Pan, Liu, and Zhang]{diao2023active}
S.~Diao, P.~Wang, Y.~Lin, R.~Pan, X.~Liu, and T.~Zhang.
\newblock Active prompting with chain-of-thought for large language models.
\newblock \emph{arXiv preprint arXiv:2302.12246}, 2023.

\bibitem[Dong et~al.(2024)Dong, Li, Dai, Zheng, Ma, Li, Xia, Xu, Wu, Chang, et~al.]{dong2024survey}
Q.~Dong, L.~Li, D.~Dai, C.~Zheng, J.~Ma, R.~Li, H.~Xia, J.~Xu, Z.~Wu, B.~Chang, et~al.
\newblock A survey on in-context learning.
\newblock In \emph{Proceedings of the 2024 Conference on Empirical Methods in Natural Language Processing}, pages 1107--1128, 2024.

\bibitem[Dosovitskiy et~al.(2020)Dosovitskiy, Beyer, Kolesnikov, Weissenborn, Zhai, Unterthiner, Dehghani, Minderer, Heigold, Gelly, et~al.]{dosovitskiy2020image}
A.~Dosovitskiy, L.~Beyer, A.~Kolesnikov, D.~Weissenborn, X.~Zhai, T.~Unterthiner, M.~Dehghani, M.~Minderer, G.~Heigold, S.~Gelly, et~al.
\newblock An image is worth 16x16 words: Transformers for image recognition at scale.
\newblock \emph{arXiv preprint arXiv:2010.11929}, 2020.

\bibitem[Fang et~al.(2025)Fang, Wang, Zhang, Zhu, Wang, Feng, and He]{fang2025reason4rec}
Y.~Fang, W.~Wang, Y.~Zhang, F.~Zhu, Q.~Wang, F.~Feng, and X.~He.
\newblock Reason4rec: Large language models for recommendation with deliberative user preference alignment.
\newblock \emph{arXiv preprint arXiv:2502.02061}, 2025.

\bibitem[Fu et~al.(2024)Fu, Wang, Wu, Chen, Huzhang, Ni, Zeng, and Zhou]{fu2024residual}
C.~Fu, K.~Wang, J.~Wu, Y.~Chen, G.~Huzhang, Y.~Ni, A.~Zeng, and Z.~Zhou.
\newblock Residual multi-task learner for applied ranking.
\newblock In \emph{Proceedings of the 30th ACM SIGKDD Conference on Knowledge Discovery and Data Mining}, pages 4974--4985, 2024.

\bibitem[Gu et~al.(2025)Gu, Zhong, Xia, Yang, Lu, Jiang, and Gai]{gu2025r4ec}
H.~Gu, R.~Zhong, Y.~Xia, W.~Yang, C.~Lu, P.~Jiang, and K.~Gai.
\newblock R4ec: A reasoning, reflection, and refinement framework for recommendation systems.
\newblock \emph{arXiv preprint arXiv:2507.17249}, 2025.

\bibitem[Hao et~al.(2024)Hao, Sukhbaatar, Su, Li, Hu, Weston, and Tian]{hao2024training}
S.~Hao, S.~Sukhbaatar, D.~Su, X.~Li, Z.~Hu, J.~Weston, and Y.~Tian.
\newblock Training large language models to reason in a continuous latent space.
\newblock \emph{arXiv preprint arXiv:2412.06769}, 2024.

\bibitem[Hu et~al.(2017)Hu, Shen, and Sun]{hu2017squeeze}
J.~Hu, L.~Shen, and G.~Sun.
\newblock Squeeze-and-excitation networks. ieee.
\newblock In \emph{CVF Conference on Computer Vision and Pattern Recognition. IEEE}, pages 7132--41, 2017.

\bibitem[Huang and Chang(2023)]{huang2023towards}
J.~Huang and K.~C.-C. Chang.
\newblock Towards reasoning in large language models: A survey.
\newblock In \emph{Findings of the Association for Computational Linguistics: ACL 2023}, pages 1049--1065, 2023.

\bibitem[Huang et~al.(2013)Huang, He, Gao, Deng, Acero, and Heck]{huang2013learning}
P.-S. Huang, X.~He, J.~Gao, L.~Deng, A.~Acero, and L.~Heck.
\newblock Learning deep structured semantic models for web search using clickthrough data.
\newblock In \emph{Proceedings of the 22nd ACM international conference on Information \& Knowledge Management}, pages 2333--2338, 2013.

\bibitem[Indyk and Motwani(1998)]{indyk1998approximate}
P.~Indyk and R.~Motwani.
\newblock Approximate nearest neighbors: towards removing the curse of dimensionality.
\newblock In \emph{Proceedings of the thirtieth annual ACM symposium on Theory of computing}, pages 604--613, 1998.

\bibitem[Isik et~al.(2024)Isik, Ponomareva, Hazimeh, Paparas, Vassilvitskii, and Koyejo]{isik2024scaling}
B.~Isik, N.~Ponomareva, H.~Hazimeh, D.~Paparas, S.~Vassilvitskii, and S.~Koyejo.
\newblock Scaling laws for downstream task performance of large language models.
\newblock In \emph{ICLR 2024 Workshop on Mathematical and Empirical Understanding of Foundation Models}, 2024.

\bibitem[Jin et~al.(2024)Jin, Xie, Zhang, Roy, Zhang, Li, Li, Tang, Wang, Meng, et~al.]{jin2024graph}
B.~Jin, C.~Xie, J.~Zhang, K.~K. Roy, Y.~Zhang, Z.~Li, R.~Li, X.~Tang, S.~Wang, Y.~Meng, et~al.
\newblock Graph chain-of-thought: Augmenting large language models by reasoning on graphs.
\newblock \emph{arXiv preprint arXiv:2404.07103}, 2024.

\bibitem[Kang and McAuley(2018)]{kang2018self}
W.-C. Kang and J.~McAuley.
\newblock Self-attentive sequential recommendation.
\newblock In \emph{2018 IEEE international conference on data mining (ICDM)}, pages 197--206. IEEE, 2018.

\bibitem[Li et~al.(2024)Li, Wang, Dai, Zhu, Yuan, Zhang, and Xia]{li2024rat}
Y.~Li, J.~Wang, T.~Dai, J.~Zhu, J.~Yuan, R.~Zhang, and S.-T. Xia.
\newblock Rat: Retrieval-augmented transformer for click-through rate prediction.
\newblock In \emph{Companion Proceedings of the ACM Web Conference 2024}, pages 867--870, 2024.

\bibitem[Lin et~al.(2024)Lin, Shan, Zhu, Du, Chen, Quan, Tang, Yu, and Zhang]{lin2024rella}
J.~Lin, R.~Shan, C.~Zhu, K.~Du, B.~Chen, S.~Quan, R.~Tang, Y.~Yu, and W.~Zhang.
\newblock Rella: Retrieval-enhanced large language models for lifelong sequential behavior comprehension in recommendation.
\newblock In \emph{Proceedings of the ACM Web Conference 2024}, pages 3497--3508, 2024.

\bibitem[Lin et~al.(2025)Lin, Dai, Xi, Liu, Chen, Zhang, Liu, Wu, Li, Zhu, et~al.]{lin2025can}
J.~Lin, X.~Dai, Y.~Xi, W.~Liu, B.~Chen, H.~Zhang, Y.~Liu, C.~Wu, X.~Li, C.~Zhu, et~al.
\newblock How can recommender systems benefit from large language models: A survey.
\newblock \emph{ACM Transactions on Information Systems}, 43\penalty0 (2):\penalty0 1--47, 2025.

\bibitem[Liu et~al.(2025)Liu, Zheng, Wang, Zhao, Wang, Chen, and Wen]{liu2025lares}
E.~Liu, B.~Zheng, X.~Wang, W.~X. Zhao, J.~Wang, S.~Chen, and J.-R. Wen.
\newblock Lares: Latent reasoning for sequential recommendation.
\newblock \emph{arXiv preprint arXiv:2505.16865}, 2025.

\bibitem[Long(2023)]{long2023large}
J.~Long.
\newblock Large language model guided tree-of-thought.
\newblock \emph{arXiv preprint arXiv:2305.08291}, 2023.

\bibitem[Madaan et~al.(2023)Madaan, Tandon, Gupta, Hallinan, Gao, Wiegreffe, Alon, Dziri, Prabhumoye, Yang, et~al.]{madaan2023self}
A.~Madaan, N.~Tandon, P.~Gupta, S.~Hallinan, L.~Gao, S.~Wiegreffe, U.~Alon, N.~Dziri, S.~Prabhumoye, Y.~Yang, et~al.
\newblock Self-refine: Iterative refinement with self-feedback.
\newblock \emph{Advances in Neural Information Processing Systems}, 36:\penalty0 46534--46594, 2023.

\bibitem[Malkov and Yashunin(2018)]{malkov2018efficient}
Y.~A. Malkov and D.~A. Yashunin.
\newblock Efficient and robust approximate nearest neighbor search using hierarchical navigable small world graphs.
\newblock \emph{IEEE transactions on pattern analysis and machine intelligence}, 42\penalty0 (4):\penalty0 824--836, 2018.

\bibitem[Marvin et~al.(2023)Marvin, Hellen, Jjingo, and Nakatumba-Nabende]{marvin2023prompt}
G.~Marvin, N.~Hellen, D.~Jjingo, and J.~Nakatumba-Nabende.
\newblock Prompt engineering in large language models.
\newblock In \emph{International conference on data intelligence and cognitive informatics}, pages 387--402. Springer, 2023.

\bibitem[Mei et~al.(2025)Mei, Yao, Ge, Wang, Bi, Cai, Liu, Li, Li, Zhang, et~al.]{mei2025survey}
L.~Mei, J.~Yao, Y.~Ge, Y.~Wang, B.~Bi, Y.~Cai, J.~Liu, M.~Li, Z.-Z. Li, D.~Zhang, et~al.
\newblock A survey of context engineering for large language models.
\newblock \emph{arXiv preprint arXiv:2507.13334}, 2025.

\bibitem[Naumov et~al.(2019)Naumov, Mudigere, Shi, Huang, Sundaraman, Park, Wang, Gupta, Wu, Azzolini, et~al.]{naumov2019deep}
M.~Naumov, D.~Mudigere, H.-J.~M. Shi, J.~Huang, N.~Sundaraman, J.~Park, X.~Wang, U.~Gupta, C.-J. Wu, A.~G. Azzolini, et~al.
\newblock Deep learning recommendation model for personalization and recommendation systems.
\newblock \emph{arXiv preprint arXiv:1906.00091}, 2019.

\bibitem[Radford et~al.(2021)Radford, Kim, Hallacy, Ramesh, Goh, Agarwal, Sastry, Askell, Mishkin, Clark, et~al.]{radford2021learning}
A.~Radford, J.~W. Kim, C.~Hallacy, A.~Ramesh, G.~Goh, S.~Agarwal, G.~Sastry, A.~Askell, P.~Mishkin, J.~Clark, et~al.
\newblock Learning transferable visual models from natural language supervision.
\newblock In \emph{International conference on machine learning}, pages 8748--8763. PmLR, 2021.

\bibitem[Ranaldi and Freitas(2024)]{ranaldi2024self}
L.~Ranaldi and A.~Freitas.
\newblock Self-refine instruction-tuning for aligning reasoning in language models.
\newblock \emph{arXiv preprint arXiv:2405.00402}, 2024.

\bibitem[Shen et~al.(2025)Shen, Yan, Zhang, Hu, Du, and He]{shen2025codi}
Z.~Shen, H.~Yan, L.~Zhang, Z.~Hu, Y.~Du, and Y.~He.
\newblock Codi: Compressing chain-of-thought into continuous space via self-distillation.
\newblock \emph{arXiv preprint arXiv:2502.21074}, 2025.

\bibitem[Su et~al.(2025)Su, Zhu, Xu, Jiao, Tian, and Zheng]{su2025token}
D.~Su, H.~Zhu, Y.~Xu, J.~Jiao, Y.~Tian, and Q.~Zheng.
\newblock Token assorted: Mixing latent and text tokens for improved language model reasoning.
\newblock \emph{arXiv preprint arXiv:2502.03275}, 2025.

\bibitem[Sun et~al.(2019)Sun, Liu, Wu, Pei, Lin, Ou, and Jiang]{sun2019bert4rec}
F.~Sun, J.~Liu, J.~Wu, C.~Pei, X.~Lin, W.~Ou, and P.~Jiang.
\newblock Bert4rec: Sequential recommendation with bidirectional encoder representations from transformer.
\newblock In \emph{Proceedings of the 28th ACM international conference on information and knowledge management}, pages 1441--1450, 2019.

\bibitem[Tang et~al.(2025)Tang, Dai, Shi, Xu, Chen, Chen, Wu, and Jiang]{tang2025think}
J.~Tang, S.~Dai, T.~Shi, J.~Xu, X.~Chen, W.~Chen, J.~Wu, and Y.~Jiang.
\newblock Think before recommend: Unleashing the latent reasoning power for sequential recommendation.
\newblock \emph{arXiv preprint arXiv:2503.22675}, 2025.

\bibitem[Vaswani et~al.(2017)Vaswani, Shazeer, Parmar, Uszkoreit, Jones, Gomez, Kaiser, and Polosukhin]{vaswani2017attention}
A.~Vaswani, N.~Shazeer, N.~Parmar, J.~Uszkoreit, L.~Jones, A.~N. Gomez, {\L}.~Kaiser, and I.~Polosukhin.
\newblock Attention is all you need.
\newblock \emph{Advances in neural information processing systems}, 30, 2017.

\bibitem[Vel{\'a}squez-Henao et~al.(2023)Vel{\'a}squez-Henao, Franco-Cardona, and Cadavid-Higuita]{velasquez2023prompt}
J.~D. Vel{\'a}squez-Henao, C.~J. Franco-Cardona, and L.~Cadavid-Higuita.
\newblock Prompt engineering: a methodology for optimizing interactions with ai-language models in the field of engineering.
\newblock \emph{Dyna}, 90\penalty0 (SPE230):\penalty0 9--17, 2023.

\bibitem[Wang et~al.(2021)Wang, Shivanna, Cheng, Jain, Lin, Hong, and Chi]{wang2021dcn}
R.~Wang, R.~Shivanna, D.~Cheng, S.~Jain, D.~Lin, L.~Hong, and E.~Chi.
\newblock Dcn v2: Improved deep \& cross network and practical lessons for web-scale learning to rank systems.
\newblock In \emph{Proceedings of the web conference 2021}, pages 1785--1797, 2021.

\bibitem[Wang et~al.(2017)Wang, Skerry-Ryan, Stanton, Wu, Weiss, Jaitly, Yang, Xiao, Chen, Bengio, et~al.]{wang2017tacotron}
Y.~Wang, R.~Skerry-Ryan, D.~Stanton, Y.~Wu, R.~J. Weiss, N.~Jaitly, Z.~Yang, Y.~Xiao, Z.~Chen, S.~Bengio, et~al.
\newblock Tacotron: Towards end-to-end speech synthesis.
\newblock \emph{arXiv preprint arXiv:1703.10135}, 2017.

\bibitem[Wang et~al.(2024)Wang, Tian, Hu, Yu, Liu, Zhang, Zhou, Pang, and Wang]{wang2024can}
Y.~Wang, C.~Tian, B.~Hu, Y.~Yu, Z.~Liu, Z.~Zhang, J.~Zhou, L.~Pang, and X.~Wang.
\newblock Can small language models be good reasoners for sequential recommendation?
\newblock In \emph{Proceedings of the ACM Web Conference 2024}, pages 3876--3887, 2024.

\bibitem[Wei et~al.(2022)Wei, Wang, Schuurmans, Bosma, Xia, Chi, Le, Zhou, et~al.]{wei2022chain}
J.~Wei, X.~Wang, D.~Schuurmans, M.~Bosma, F.~Xia, E.~Chi, Q.~V. Le, D.~Zhou, et~al.
\newblock Chain-of-thought prompting elicits reasoning in large language models.
\newblock \emph{Advances in neural information processing systems}, 35:\penalty0 24824--24837, 2022.

\bibitem[Wu et~al.(2024)Wu, Zheng, Qiu, Wang, Gu, Shen, Qin, Zhu, Zhu, Liu, et~al.]{wu2024survey}
L.~Wu, Z.~Zheng, Z.~Qiu, H.~Wang, H.~Gu, T.~Shen, C.~Qin, C.~Zhu, H.~Zhu, Q.~Liu, et~al.
\newblock A survey on large language models for recommendation.
\newblock \emph{World Wide Web}, 27\penalty0 (5):\penalty0 60, 2024.

\bibitem[Yan et~al.(2024)Yan, Zhu, Wang, Gui, and He]{yan2024mirror}
H.~Yan, Q.~Zhu, X.~Wang, L.~Gui, and Y.~He.
\newblock Mirror: A multiple-perspective self-reflection method for knowledge-rich reasoning.
\newblock \emph{arXiv preprint arXiv:2402.14963}, 2024.

\bibitem[Yao et~al.(2023{\natexlab{a}})Yao, Yu, Zhao, Shafran, Griffiths, Cao, and Narasimhan]{yao2023tree}
S.~Yao, D.~Yu, J.~Zhao, I.~Shafran, T.~Griffiths, Y.~Cao, and K.~Narasimhan.
\newblock Tree of thoughts: Deliberate problem solving with large language models.
\newblock \emph{Advances in neural information processing systems}, 36:\penalty0 11809--11822, 2023{\natexlab{a}}.

\bibitem[Yao et~al.(2023{\natexlab{b}})Yao, Li, and Zhao]{yao2023beyond}
Y.~Yao, Z.~Li, and H.~Zhao.
\newblock Beyond chain-of-thought, effective graph-of-thought reasoning in language models.
\newblock \emph{arXiv preprint arXiv:2305.16582}, 2023{\natexlab{b}}.

\bibitem[Yu et~al.(2025)Yu, Fu, Zhang, Lv, Wu, and Wu]{yu2025thinkrec}
Q.~Yu, K.~Fu, S.~Zhang, Z.~Lv, F.~Wu, and F.~Wu.
\newblock Thinkrec: Thinking-based recommendation via llm.
\newblock \emph{arXiv preprint arXiv:2505.15091}, 2025.

\bibitem[Zhai et~al.(2024)Zhai, Liao, Liu, Wang, Li, Cao, Gao, Gong, Gu, He, et~al.]{zhai2024actions}
J.~Zhai, L.~Liao, X.~Liu, Y.~Wang, R.~Li, X.~Cao, L.~Gao, Z.~Gong, F.~Gu, J.~He, et~al.
\newblock Actions speak louder than words: trillion-parameter sequential transducers for generative recommendations.
\newblock In \emph{Proceedings of the 41st International Conference on Machine Learning}, pages 58484--58509, 2024.

\bibitem[Zhang et~al.(2025{\natexlab{a}})Zhang, Zhang, Sun, Lu, Zhao, Chen, and Wen]{zhang2025slow}
J.~Zhang, B.~Zhang, W.~Sun, H.~Lu, W.~X. Zhao, Y.~Chen, and J.-R. Wen.
\newblock Slow thinking for sequential recommendation.
\newblock \emph{arXiv preprint arXiv:2504.09627}, 2025{\natexlab{a}}.

\bibitem[Zhang et~al.(2025{\natexlab{b}})Zhang, Xu, Zhao, Wang, Feng, He, and Chua]{zhang2025reinforced}
Y.~Zhang, W.~Xu, X.~Zhao, W.~Wang, F.~Feng, X.~He, and T.-S. Chua.
\newblock Reinforced latent reasoning for llm-based recommendation.
\newblock \emph{arXiv preprint arXiv:2505.19092}, 2025{\natexlab{b}}.

\bibitem[Zhang et~al.(2022)Zhang, Zhang, Li, and Smola]{zhang2022automatic}
Z.~Zhang, A.~Zhang, M.~Li, and A.~Smola.
\newblock Automatic chain of thought prompting in large language models.
\newblock \emph{arXiv preprint arXiv:2210.03493}, 2022.

\bibitem[Zhao et~al.(2023)Zhao, Zhou, Li, Tang, Wang, Hou, Min, Zhang, Zhang, Dong, et~al.]{zhao2023survey}
W.~X. Zhao, K.~Zhou, J.~Li, T.~Tang, X.~Wang, Y.~Hou, Y.~Min, B.~Zhang, J.~Zhang, Z.~Dong, et~al.
\newblock A survey of large language models.
\newblock \emph{arXiv preprint arXiv:2303.18223}, 1\penalty0 (2), 2023.

\bibitem[Zhou et~al.(2018)Zhou, Zhu, Song, Fan, Zhu, Ma, Yan, Jin, Li, and Gai]{zhou2018deep}
G.~Zhou, X.~Zhu, C.~Song, Y.~Fan, H.~Zhu, X.~Ma, Y.~Yan, J.~Jin, H.~Li, and K.~Gai.
\newblock Deep interest network for click-through rate prediction.
\newblock In \emph{Proceedings of the 24th ACM SIGKDD international conference on knowledge discovery \& data mining}, pages 1059--1068, 2018.

\bibitem[Zhu et~al.(2025)Zhu, Peng, Cheng, Qu, Huang, Zhu, Wang, Xue, Zhang, Shan, et~al.]{zhu2025survey}
R.-J. Zhu, T.~Peng, T.~Cheng, X.~Qu, J.~Huang, D.~Zhu, H.~Wang, K.~Xue, X.~Zhang, Y.~Shan, et~al.
\newblock A survey on latent reasoning.
\newblock \emph{arXiv preprint arXiv:2507.06203}, 2025.

\bibitem[Zhu et~al.(2023)Zhu, Yuan, Wang, Liu, Liu, Deng, Chen, Liu, Dou, and Wen]{zhu2023large}
Y.~Zhu, H.~Yuan, S.~Wang, J.~Liu, W.~Liu, C.~Deng, H.~Chen, Z.~Liu, Z.~Dou, and J.-R. Wen.
\newblock Large language models for information retrieval: A survey.
\newblock \emph{arXiv preprint arXiv:2308.07107}, 2023.

\end{thebibliography}
